\documentclass[electronic]{vgtc}             % electronic version

\ifpdf%                                % if we use pdflatex
  \pdfoutput=1\relax                   % create PDFs from pdfLaTeX
  \usepackage{graphicx}                % allow us to embed graphics files
  \DeclareGraphicsExtensions{.pdf,.png,.jpg,.jpeg} % for pdflatex we expect .pdf, .png, or .jpg files
\else%                                 % else we use pure latex
  \usepackage{graphicx}                % allow us to embed graphics files
  \DeclareGraphicsExtensions{.eps}     % for pure latex we expect eps files
\fi%

\graphicspath{{figures/}{pictures/}{images/}{./}} % where to search for the images

\usepackage{microtype}                 % use micro-typography (slightly more compact, better to read)
\PassOptionsToPackage{warn}{textcomp}  % to address font issues with \textrightarrow
\usepackage{textcomp}                  % use better special symbols
\usepackage{mathptmx}                  % use matching math font
\usepackage{times}                     % we use Times as the main font
\usepackage{cite}                      % needed to automatically sort the references
\usepackage{tabu}                      % only used for the table example
\usepackage{booktabs}                  % only used for the table example
\usepackage{verbatim}
\usepackage{makecell}
\usepackage{multirow}
\usepackage{amsmath}

\newcommand{\tabincell}[2]{\begin{tabular}{@{}#1@{}}#2\end{tabular}}

\newcommand{\redsout}[1]{}
\usepackage[ruled,linesnumbered]{algorithm2e}   % algorithm writing

\onlineid{0}

\vgtccategory{Research}

\vgtcinsertpkg

\title{Label Guidance based Object Locating in Virtual Reality}
\author
{
\vspace{-0.6cm}
     Xiaoheng Wei\textsuperscript{1} \\
    \and Xuehuai Shi\textsuperscript{1}\\
    \and Lili Wang\textsuperscript{1;2;3}
    \thanks{Lili Wang is the corresponding author: wanglily@buaa.edu.cn}\\
}
\affiliation{\scriptsize\textsuperscript{1}State Key Laboratory of Virtual Reality Technology and Systems, Beihang University, Beijing, China\\\textsuperscript{2}Peng Cheng Laboratory, Shengzhen, China\\
\textsuperscript{3}Beijing Advanced Innovation Center for Biomedical Engineering, Beihang University, Beijing, China.\\}

\teaser{
  \centering
  \includegraphics[width=\linewidth]{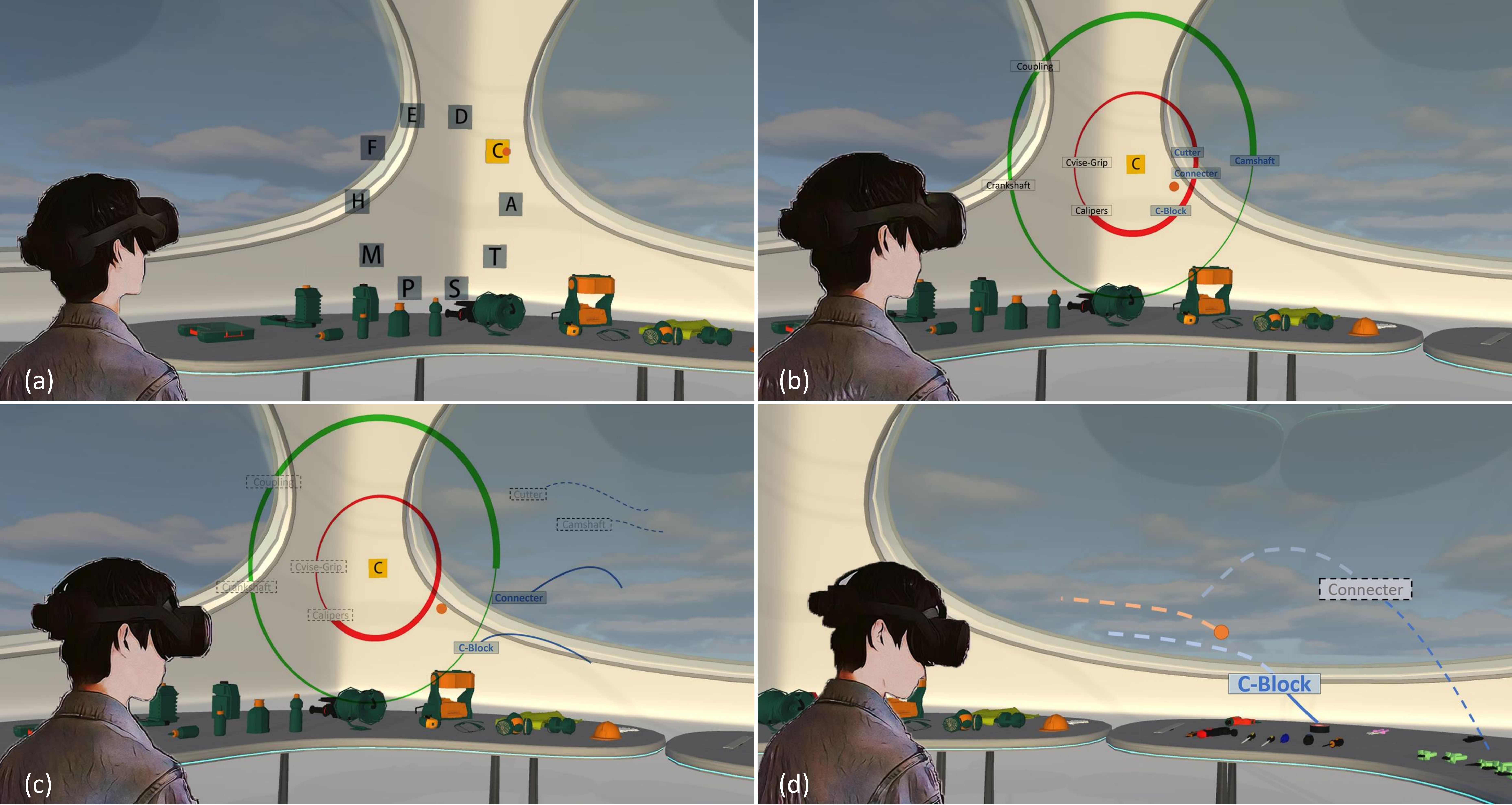}
  \caption{A user uses the two-level hierarchical sorted and orientated label layout to locate the target object. Firstly, the user selects the initial letter on the first-level sorted circle layout (a). Secondly, the user selects the candidate labels on the second-level sorted circle layout (b). Thirdly, the candidate labels are updated according to the movement of the gaze point (c). At last, {\redsout{the target label guides}} the user {\redsout{to locate}}{{locates}} the { target} object (d).}
  \label{fig:teaser}
}
\abstract{%
Object locating in virtual reality (VR) {\redsout{is}has been} widely used in many VR applications, such as virtual assembly, virtual repair, virtual remote coaching{\redsout{, and so on}}.
However, when there are a large number of objects in the virtual environment{(VE)}, the user cannot locate the target object efficiently and comfortably. 
In this paper, we propose a label guidance based object locating method for {\redsout{localizing}}{locating} the target object efficiently in VR. 
Firstly, we introduce the label guidance based object locating pipeline {to improve the efficiency of the object locating}. 
It arranges the labels of all objects on the same screen, lets the user select the target {\redsout{label}labels} first, and {then} uses the flying labels to guide the user to the target object. 
Then we summarize five principles for constructing the label layout for object locating and propose a two-level hierarchical sorted and orientated label layout based on the five principles for the user to select the candidate labels efficiently and comfortably. 
After that, we propose the view and gaze based label guidance method for guiding the user to locate the target object based on the selected candidate labels. 
It generates specific flying trajectories for candidate labels, updates the flying speed of candidate labels, keeps valid candidate labels , and removes the invalid candidate labels in real time during {\redsout{the process of}} object locating with the guidance of the candidate labels. 
Compared with the traditional method, the user study results show that {\redsout{using}} our method significantly improves efficiency and reduces task load for object locating.
} % end of abstract

\CCScatlist{
    {Virtual Reality}, {Object Locating}, {Label Guidance}
}

\begin{document}

%% The ``\maketitle'' command must be the first command after the
%% ``\begin{document}'' command. It prepares and prints the title block.

%% the only exception to this rule is the \firstsection command
\firstsection{Introduction}
\maketitle
Recently, {\redsout{virtual reality (VR)}{VR}} technology has made {\redsout{great}}{significant} progress and has been applied to many industries such as manufacturing, entertainment {\redsout{industry}}, {and} education {\redsout{industry, and so on}}.
{\redsout{In these applications, especially in multi-person collaborative operation applications such as virtual assemble, virtual repair, virtual remote coaching, etc., object locating is widely used}}{Object locating is widely used\cite{bodonyi2020efficient}, especially in multi-person collaborative operation applications such as virtual assembly, virtual repair, and virtual remote coaching.}
However, inefficient object locating methods will greatly reduce the users' experience of these applications.

% 删掉的东西
% {For example, in a virtual collaborative machine assembly application, when the remote supervisor asks the local {\redsout{worker}}{workers} to assemble a specific part to a machine from a large number of local parts, the local workers cannot quickly and accurately assemble the specific part to the machine, which will {\redsout{greatly}}{significantly} reduce the efficiency of the virtual collaborative machine assembly.}

% 在虚拟环境中，给物体打上标签有利于用户快速locate object，但是当场景中的物体数量变得很多时，标签和物体之间相互遮挡反而会降低locating object的效率，给用户带来困扰。This
% - challenge is locate the object efficiently and comfortably when the number of objects is huge \\
% - VR Locating object, state of art , the problem\\

Locating the target object with the label guidance is an idea for efficiently object locating in VR.
However, it brings three challenges.
In the {\redsout{virtual environment}}VE with a large number of labeled objects, the user has to search for the target in all directions.
If {\redsout{the label of the target object is blocked by other objects}}{other objects block the label of the target object}, the user also needs to walk around to find it, which makes it more difficult for the user to find the target object.
So the first challenge is to design the pipeline of label guidance for locating the target object efficiently and comfortably.
In the VE that contains a large number of labeled objects, the user can not find the candidate labels quickly.
So the second challenge is to construct the sorted label layout so {\redsout{that}}the user can select the candidate labels efficiently.
When using the candidate labels to guide the user to locate the target object, if the candidate labels are too close or too far away from the user’s position, {\redsout{or}}the moving speed of the candidate labels is too fast or slow, or the invalid candidate labels are not removed in time, the user can not locate the target object accurately and comfortably.
So the third challenge is to guide the user to locate the target object accurately and comfortably.

% the label guidance based object locating method
% the two-level hierarchical sorted and orientated label layout
% the view and gaze based label guidance method
In this paper, we propose a label guidance based object locating method to improve the efficiency of locating the target object in VR applications.
For the first challenge, we propose a label guidance based object locating pipeline, which arranges the labels of all objects on the same screen, lets the user select the target label first, and uses the flying labels to guide the user to the target object.
In order to let the user select the candidate labels efficiently and comfortably, {\redsout{five principles for constructing the label layout for object locating are introduced}}{we summarize five principles for constructing the label layout for object locating}. 
For the second challenge, we design and construct the two-level hierarchical sorted and orientated label layout based on five summarized principles.
The user selects the candidate labels in the two-level hierarchical sorted and orientated label layout.
For the third challenge, {\redsout{the view and gaze based label guidance method is proposed to generate specific flying trajectories for candidate labels}}{we propose the view and gaze based label guidance method to generate specific flying trajectories for candidate labels}, update the flying speed of candidate labels, keep valid candidate labels and remove the invalid {\redsout{candidate labels}}{ones} in {real time} during the process of object locating.
Finally, the user uses the valid flying candidate label to locate the target object.
We design a user study of two tasks to evaluate the performance of our method.
Compared with the traditional method, the results show that our method significantly improves efficiency and reduces task load for object locating.
Figure \ref{fig:teaser} shows the process of a user locating an object by our method.

% the label guidance based object locating method
% the two-level hierarchical sorted and orientated label layout
% the view and gaze based label guidance method
In summary, the contributions of our method are as follows: 
\emph{1)} we propose a label guidance based object locating pipeline to improve the efficiency of locating the target object in VR applications; 
\emph{2)} we introduce a two-level hierarchical sorted and orientated label layout to provide sort and orientation cues of the target object; 
\emph{3)} we introduce a view and gaze based label guidance method to optimize the label guidance path, update the guidance speed of labels, keep valid labels and remove the invalid labels in {real time} during the process of object locating; 
\emph{4)} we design a user study to evaluate the efficiency of our method.

% %% \section{Introduction} %for journal use above \firstsection{..} instead
% This template is for papers of VGTC-sponsored conferences which are \emph{\emph{not}} published in a special issue of TVCG.

\vspace{-0.3em}
\section{Related work}

% \subsection{Object Locating in VR}
% - object locating: visual search, object locate
In this section, we {\redsout{give a brief review of}}{briefly review} the previous work on the label layout and out-of-view guidance related to our method.

\vspace{-0.3em}
\subsection{Label Layout}

% 插画，二维图像 的标签布局
Label layout is widely studied in two-dimensional and three-dimensional images.
Fink et al.\cite{fink2012algorithms} used a circle and cluster layout to place labels on the focus regions of the two-dimensional map.
Heinsohn et al.\cite{heinsohn2014boundary} placed labels for the dynamic nature of the focus region , and users can obtain details on their demand through the cluster layout during {the} overview. 
Kouvril et al.\cite{kouvril2018labels} extracted the hierarchical structure of the objects and {\redsout{labels}}{labeled} different levels of objects to deal with large hierarchical environments.

% VR/AR/全景视频 的标签布局
Tatzgern et al.\cite{tatzgern2014hedgehog} constrained the placement of labels in 3D object space according to user viewpoint transformation.
Cmolik et al.\cite{cmolik2020mixed} proposed a hybrid label layout, which determines the placement position and type of labels according to the threshold specified by the user. 
Zhou et al.\cite{zhou2021partially} determined a 2D label layout plane by user view direction and arranged the labels on the circle of the 2D plane in alphabetical order.
% 删掉的东西
% {The commonality of their application was that the virtual labels were distributed around an in-view physical object through leaders and anchors.} 
Grasset et al.\cite{grasset2012image} summarized the basic rules and criteria for placing labels in VR scenes , and they proposed an image-based approach to identify geometric constraints for placing labels.
Mcnamara et al.\cite{mcnamara2019information} used eye tracking to calculate the user's potential object of interest and adjusted the placement strategy of label information in the complex VE. 
Jia et al.\cite{jia2021semantic} established the label placement constraints according to the user's semantic perception of the image.

The above three label placement methods all had objects scattered in multiple locations of a scene.
Generally, {\redsout{whether it is for an in-view object or for multiple objects of a scene, label layout requires that labels cannot occlude each other}}{label layout requires that labels cannot occlude each other, whether for an in-view object or multiple objects of a scene}.
The position of the label should be adjusted according to the user's viewpoint and {\redsout{should}} always remain in the user's current view, and the distance between the label and its anchor object should be as close as possible.
Our method provides the label layout that remains in the user's current view, {\redsout{and has}}{with} no occlusion among labels, and {\redsout{has}} good interactivity.
% Lin et al.\cite{lin2021labeling} designed space of labels based the angular direction of the object in AR application for situated visual search.

% 大量标签下的 标签布局
When the scene has too many labels, the display of the user's current view quickly becomes cluttered. 
Zhang et al.\cite{zhang2010annotating} reduced the number of labels displayed by scoring building importance and scheduling annotations in the video.
Tatzgern et al.\cite{tatzgern2013dynamic} created a temporally coherent layout for compact label annotations, thus avoiding visual interference when the viewpoint changes. 
Then they\cite{tatzgern2016adaptive} created an information hierarchy to cluster a large number of labels by weighted calculation of user-defined spatial and non-spatial attributes. The above methods adaptively adjusted the information density by reducing the labels displayed on the screen by filtering or clustering operations.
We also use this idea for the object locating task, cluster the labels according to their initial letter, and then expand the labels according to the selected initial letter.

% - label layout  method 3.2: label layout in AR, view management
% 我们的总结 
% {our method}
\vspace{-0.3em}
\subsection{Out-of-View Object Guidance}

The guidance technology of out-of-view objects is divided into three types according to visualization methods. 
One is to visualize out-of-view objects as abstract visualized symbols and encode the information of out-of-view objects into the attributes of the visualized symbols.
Peterson\cite{peterson2008managing} reduced the label visual cluster by using stereoscopic discrimination.
Schwerdtfeer\cite{schwerdtfeger2008supporting} compared the frame, tunnel, and arrow visualization guidance.
Renner\cite{renner2017attention} evaluated AR-based guiding techniques based on images, funnel, arrow and proposed a SWAVE guidance technique based on eye gaze information.
EyeSee360\cite{gruenefeld2017eyesee360} is a 2D visualization technique with distance-encoding and direction-encoding.
The flyingarrow\cite{gruenefeld2018flyingarrow} flew to the location of the out-of-view object according to the user's current sight.
Bork\cite{bork2018towards} proposed a mirror ball of all virtual objects' reflections to provide positive hints and a 3D {\redsout{rader}}{radar} method for visualizing user positions and {\redsout{out of view}}{out-of-view} objects.

% 一种是用声音等信号，这个简略写
The other is to use non-visual cues, such as vibro-tactile cues\cite{lehtinen2012dynamic, lindeman2003effective}, auditory cues\cite{van2008pip, mcintire2010visual} or blend several cues\cite{marquardt2020comparing} {\redsout{and so on}}.

Another guidance technology is labeling. The advantage of labeling is to support bidirectional retrieval workflows for object-to-label and label-to-object lookups\cite{lin2021labeling}. 
Kruijff\cite{kruijff2018influence} evaluated the impact of virtual label characteristics such as color, size, and leader lines on the search performance and gave suggestions on label design in wide FOV augmented reality displays.
Lin\cite{lin2021labeling} explored the design space of labels in {\redsout{AR application for situated visual search and compared three representative AR labeling techniques which respectively encode the different information of objects}}{AR applications for situated visual search and compared three representative AR labeling techniques that encode different objects' different information}. They demonstrated that angle-encoded labels with directional cues perform best, and our method also uses the labels to guide the user to the target object.

% 总结我们的方法和related work的关系。

% visual search

% 每小节 8-10篇文章，分2-3类，按年份分别讲，20年以上，每小节最后一段话总结

% the label guidance based object locating method
% the two-level hierarchical sorted and orientated label layout
% the view and gaze based label guidance method
\vspace{-0.3em}
\section{Method}

In order to locate the target object in the VE with a large number of objects more efficiently and comfortably, we propose a label guidance based object locating method.
In this section, {\redsout{first}}we {first} describe our label guidance based object locating pipeline in Section \ref{LabelGuidance}. Then we introduce the two-level hierarchical sorted and orientated label layout to provide orientation cues {\redsout{of}}{for} the target object in Section \ref{LabelLayout}. At last, we introduce the view and gaze based label guidance method to optimize the guidance path, update the guidance speed of labels, keep valid labels and remove the invalid labels in {real time} during the process of object locating in Section \ref{LabelPath}.

% In order to find out-of-view objects more breezily and efficiently, we propose a clustering selection scheme based on  initials dictionary order. We will describe the specific selection pipeline in Section 3.1. Then, we will describe the label layout algorithm based on direction indication of the objects and lexicographical order information in Section 3.2 and the out-of-view labels auxiliary tracking algorithm for label flying orbit and speed optimization in Section 3.3.

\vspace{-0.3em}
\subsection{Label Guidance based Object Locating Pipeline} \label{LabelGuidance}

{\redsout{When a user stands in the VE that has a large number of objects with labels around him, it is not easy for the user to locate the target object with a certain label.}}{When a user stands in the VE with many objects around him, it is not easy for the user to locate the target object with a specific label.} The user has to turn around and look in all directions to search for the target object. 
% 删掉的东西
% {If {\redsout{the label of the target object is occluded by others}}{others occlude the label of the target object}, the user needs to walk around to find it. }
In order to improve the efficiency of object locating, the idea of our method is to arrange the labels of all objects on the same screen, let the user select the candidate labels first, and use the candidate labels to guide the user to the target {object}. The pipeline of our label guidance based object locating has the following steps.

% a hierarchical circle of the labels
% The hierarchical label circle
% a two-level hierarchical sorted and orientated label layout

\textbf{Firstly}, we construct the two-level hierarchical sorted and orientated label layout.
% a two-level hierarchical sorted and orientated label layout is constructed and showed on the screen when the user requires. 
The two-level hierarchical sorted and orientated label layout contains the first-level sorted circle layout and the second-level sorted circle layout.
The first-level sorted circle layout is a circle displayed in the user's view.
The initial letters of labels are arranged on the edge of the first-level sorted circle layout in {\redsout{the}} counterclockwise alphabetical order.
The second-level sorted circle layout appears in the user's view when the first-level sorted circle layout disappears.
It is a concentric circle, and
the sorted labels with orientation cues are arranged on the edge of all circles in the second-level sorted circle layout.

% 硬伤
\textbf{Secondly}, 
the user selects the target label {with head movement} on the two-level hierarchical sorted and orientated label layout{\redsout{with head movement}}. 
% rebutal增加解释
We refer {to} the center of the user view as the gaze point since {\redsout{we didn't use a gaze capture device to capture the gaze point of the user}}{the use of gaze capture device will bring two problems.
First, the gaze point position may be located in the current view's edge area, which will break the in-view principle in subsection 3.2.
Second, since the user's gaze position and gaze direction may jump continuously over a while, if the location and the normal direction of the label layout change according to the user's gaze position and gaze direction frame by frame precisely, the label layout will jump and flicker on the screen, which will cause motion sickness}.

When the user needs to select the label, the first-level sorted circle layout is displayed on the screen.
Then the user performs the initial letter selection.
In the initial letter selection,
the user uses gaze based dwell-time method \cite{xu2019ringtext} to select the initial letter, i.e., the user's gaze stays on the initial letter for 400\emph{ms}.
After that, the first-level sorted circle layout disappears, and the second-level sorted circle layout unfolds, on which the labels with the selected initial letter are shown. 
The user selects the candidate labels on {\redsout{the}}the second-level sorted circle layout.
In the candidate labels selection,
we first calculate the orientation of each label{\redsout{, which is calculated}} from the label position and the center position of the second-level sorted circle layout. 
% we first calculate the orientation of each label from the label position and the center position of the second-level circle.
Those labels whose orientations are less than 90$^\circ$ from the moving direction of the gaze point are regarded as candidate labels.
Then we record the user's gaze point in each frame, compute the moving direction of the gaze point by the least-squares fitting function, and the gaze moving direction is used to select the candidate labels.

% The orientation of each label is calculated from the label position and the center position of the second-level circle. 
% % Those labels whose orientation differ from the gaze point moving direction by less than 90 degrees are selected as candidate labels.
% We regard those labels whose orientation are less than 90$^\circ$ from the moving direction of the gaze point as candidate labels.

\textbf{Thirdly}, the candidate labels guide the user to locate the target object by flying in a specific trajectory.
The candidate labels fly back to their anchor objects according to the optimized paths to avoid labels penetrating the user's body during the guidance process.

\textbf{Simultaneously}, the user locates the target object with the guidance of candidate labels.
The user moves his head and uses his gaze to follow and select the target label from the flying candidate labels. The moving direction of the candidate labels and the moving direction of the user's gaze {\redsout{are}}{{is}} calculated in real time. If the moving direction of the candidate deviates from the user's gaze moving direction by more than 90$^\circ$ during the flight, it will be regarded as the invalid candidate label and be removed. 
% The user locates the target with the left label fly to it.
% The user locates the target object with the left flying candidate labels.
The user uses the valid flying candidate labels to locate the target object.

\subsection{Two-level Hierarchical Sorted and Orientated Label Layout}\label{LabelLayout}

Previous work {\redsout{of}}{on} labeling objects in VR and AR has introduced some basic principles of label layout, such as no occlusion among labels, closer distance between the label and its anchor object, and excellent interactivity. 
In the label guidance based object locating task, in order to search the labels and locate the corresponding objects efficiently, {\redsout{We}}{we} summarize five principles.

\textbf{Principle 1 $In\ View.$}
% \textbf{1) Visibility.}
There are many objects in the VE around the user. When the user needs to locate the target object, she/he needs to look around constantly with the traditional method. The idea of the label-guided method is to find the label of the target first and then guide the user to locate the target through the movement of the label.
In order to find the label efficiently, the initial position of the label is preferably within the user's view, so that the user does not need to look around. After the label is found, if the 
anchor object is not in the current view, the label will fly and guide the user to turn his head and transform to the view containing the anchor object.

% \textbf{2) Sorted.} 
% Regularity 
\textbf{Principle 2 $Sorted\ Label.$} 
To further speed up the label selection, the label arrangement needs to be organized in an orderly manner. Usually, the user is most familiar with the alphabetical order {\redsout{, so}}{so that} the labels can be arranged in that order.
Due to a large number of objects, the speed of label selection is still limited if all labels are arranged in alphabetical order. 
Given that the user is very familiar with the first letter index of the dictionary and the way words are arranged alphabetically under the same first letter, it is {\redsout{a good}}{an excellent} choice to use a hierarchical method for label layout arrangement.

% Labels are placed on multiple circles whose center are the current gaze point. The labels on the same circle are arranged clockwise or counterclockwise in lexicographical order. The inner circles with smaller radius are arranged first to ensure minimum numbers of circle count. The circle visualization scheme also encodes the sequence information, such as the circle line thickness is aligned with the lexicographical sequence, and the circle line color is aligned with the clockwise direction.

\textbf{Principle 3 $Interaction.$} 
The label layout allows users to make convenient and robust label selection with only minor movements, combined with the steering to track labels and locate anchor objects. There are no restrictions on the specific interaction method. That is to say, {\redsout{users can use the handle to select, or you can use the hand-free mode, such as head movement, to select}}{users can use the handle or hand-free mode to select labels.}

\textbf{Principle 4 $Orientation.$}
The display position of the label on the screen can encode the orientation information of the object indicated by the label, helping the user obtain the relative position of the label and the anchor object, understand the potential moving direction of the flying label and {\redsout{easily}}{effortlessly} follow the label guidance.

\textbf{Principle 5 $Distance.$}
If the orientation cue of the anchor object contained in the label is a specific value, this constraint is too strict because the label position should reflect orientation information about its anchor object and not be strictly limited to a precise direction value. So a range of orientations is usually used. {\redsout{Then the size of this range can be determined according to the distance between the label and its anchor object. The greater the distance, the greater the range, and vice versa.}}{Range size should be determined by the distance between the label and its anchor object, and the greater the distance, the greater the range.}

% The label layout principle requires keeping the label-object distance as closer as possible.  We adjust the label position of circle to get a global approximate optimal solution. If the label is farther away from the anchor object, the range of adjustable positions in circle will be greater. In this way, the label at the top left of the screen can not fly back to the bottom right, i.e. the label is always displayed near the object.

%  all labels of the objects are displayed in the user view when user requires the label guidance by pressing the button on the handle. 
According to the five principles above, we propose the two-level hierarchical sorted and orientated label layout. 
Based on \textbf{Principle 1},
all labels can be displayed in the user view by pressing the button on the handle when the user needs the label guidance.

Based on \textbf{Principle 2} and \textbf{Principle 3}, 
we design a two-level hierarchical sorted and orientated label layout, which contains the first-level sorted circle layout and the second-level sorted circle layout.
{\redsout{The first-level sorted circle layout is displayed first, which is a circle displayed in the user's view.}}{The first-level sorted circle layout is a circle displayed in the user's view.}
The initial letters of the labels are arranged in counterclockwise alphabetical order from the three o'clock position in the first-level sorted circle layout.
The first-level sorted circle layout is placed in the center of the user's view, and its radius is calculated using the central vision of the human field of view (FOV) \cite{trepkowski2019effect}.
After the initial letter selection, the first-level sorted circle layout disappears.
The second-level sorted circle layout with the sorted label for the selected initial letter unfolds. 
The second-level sorted circle layout is a concentric circle displayed in the user's view.
The sorted labels are placed on each circle of the concentric circle.
The second-level sorted circle layout 
centers on the selected initial letter and uses the central vision of FOV to compute its radius.

Based on \textbf{Principle 4} and \textbf{Principle 5}, the labels on the second-level sorted circle layout are placed to indicate the orientation of their anchor objects in the VE.

Given the label set $L$, the anchor object set $O$ of $L$, the user gaze point $G$, and the user view direction $d$, the maximum number of circles $N$ in the second-level sorted circle layout $MCL$, the maximum number of iterations for relaxation $\#N_{it}$, the second-level sorted circle layout $MCL$ is calculated by Algorithm \ref{alg_layout}.
\vspace{-1.0em} 
\begin{algorithm}
\caption{Second-level Sorted Circle Layout}
\label{alg_layout}
\KwData{label set $L$, object set $O$, user gaze point $G$, user view direction $d$, maximum number of circles $N$, maximum number of iterations for relaxation $\#N_{it}$}
\KwResult{second-level sorted circle layout $MCL$}
$ \Pi \leftarrow $plane($G$, $d$) \;
$SCL \leftarrow  initArray()$\;
\For {$l_i \in L$}{
    % $GO_i \leftarrow G - l_i.pos$\;
    $\overrightarrow{v} \leftarrow G-o_i$\;
    $\overrightarrow{v_{\Pi}} \leftarrow $ project($\overrightarrow{v}, \Pi$)\;
    $l_i.dis \leftarrow |\overrightarrow{v_{\Pi}}|$\;
    $l_i.rad \leftarrow$ radian$(\overrightarrow{v_{\Pi}})$\;
    $l_i.rad_p \leftarrow$ $l_i.rad$\;
    $l_i.ran \leftarrow $range$(l_i)$ \;
    % $v_i  \leftarrow  initVector(|v|, radian(v))$\;
    $SCL \leftarrow SCL + l_i$\;
    % $v_i \leftarrow $Project($GO_i$, $\Pi$) \;
    % $SCL.\{$Angle($d_i$)\} $\leftarrow l_i$\;
    % $SCL.\{$$d_i.angle$\} $\leftarrow l_i$\;
}
$MCL \leftarrow initArray(N)$\;
$k=0$\;
\While{$SCL \neq \emptyset$}{
    $MCL[k]$, $SCL$ $\leftarrow$ maxSortedSubseq ($MCL[k]$, $SCL$)\;
    % $MCL[k]$, $SCL$ \leftarrow$ MaxSortedSubsequence ($MCL[k]$, $SCL$)\;
    % \For{$l_i \in SCL$}{$MCL[k]$ $\leftarrow$ Insert($l_i$, $MCL[k]$)\;}
    \For{$i \in [0,len(SCL)]$}
    {$MCL[k]$, $SCL$ $\leftarrow$ insert($MCL[k]$, $SCL$, $SCL[i]$)\;}
    
    $MCL[k] \leftarrow $ relax($MCL[k]$, $k$, $\#N_{it}$)\;
    $k \leftarrow k + 1$\;
}
\end{algorithm}
\vspace{-1.5em} 

Each label $l$ in $L$ has four attributes: $dis$, $rad_p$, $rad$, and $ran$.
$l.dis$ is the projection length in screen space of the distance between the anchor object's position of $l$ and gaze point position in the VE.
Since labels are placed on $MCL$, we only need to record the label's radian to determine the position of the label on $MCL$.
$l.rad_p$ is the initial radian of $l$ in $MCL$.
$l.rad$ is the current radian of $l$ in $MCL$.
$l.ran$ is the sliding range of $l$ in $MCL$, which is an interval [$l.ran.ran_{min}$, $l.ran.ran_{max}$] formed by $l.ran.ran_{min}$ and $l.ran.ran_{max}$.

\begin{figure}[h]
  \vspace{-0.5em}
 \centering % avoid the use of \begin{center}...\end{center} and use \centering instead (more compact)
 \includegraphics[width=0.5\columnwidth]{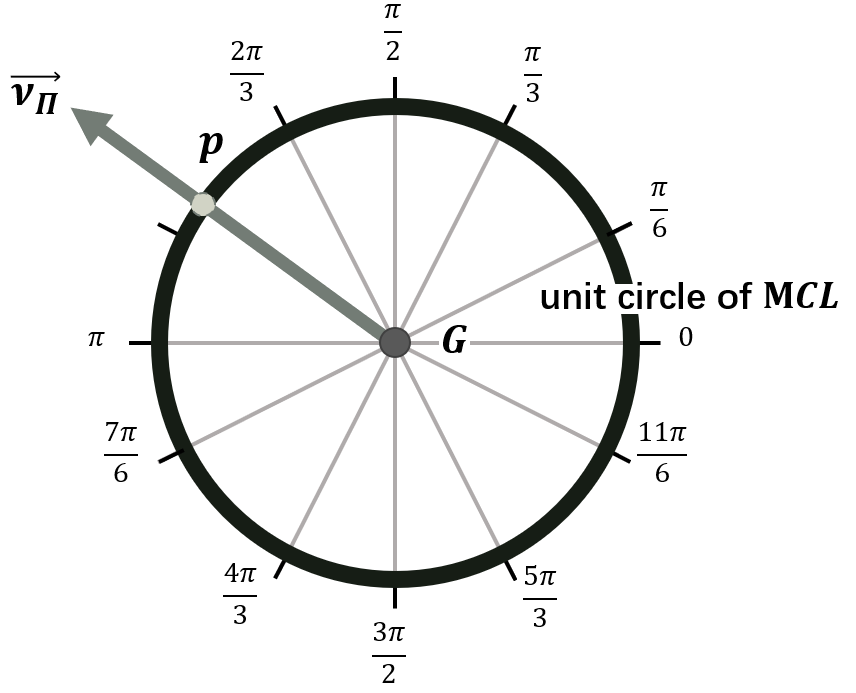}
 \caption{Computation of the label's radian.}
 \label{fig:radianCal}
  \vspace{-1.0em}
\end{figure}

Firstly we use the gaze point $G$ as the center point, and the user's view direction $d$ as the normal vector to generate the layout plane $\Pi$ (line 1).
The single circle layout $SCL$ is initialized in line 2.
We initialize the attributes of each label $l$ in the label set $L$ and store them in $SCL$ (lines 3-11).
For each label $l_i$ in $L$ (line 3), 
we get the vector $\overrightarrow{v}$ from $l_i$'s anchor object position $o_i$ to the gaze point $G$ (line 4).
Then we project $\overrightarrow{v}$ on $\Pi$ to get the vector $\overrightarrow{v_{\Pi}}$ (line 5).
After that, we initialize the attributes of $l_i$. 
$l_i.dis$ is initialized as the modulus length of $\overrightarrow{v_{\Pi}}$ (line 6).
$l_i.rad$ is calculated by $radian$ function (line 7).
As shown in Figure \ref{fig:radianCal}, we get the intersection $p$ of $\overrightarrow{v_{\Pi}}$ on the edge of the unit circle in the screen space, and $l_i.rad$ is set as the radian of $p$ on the edge of the unit circle.
$l_i.rad_p$ is set the same as $l_i.rad$ (lines 8).
$l_i.ran$ is the range [$ran_{min}, ran_{max}$], where $ran_{min}, ran_{max}$ are calculated by Equation \ref{range_2} (line 9). After the attributes of $l_i$ are initialized, we add $l_i$ to $SCL$ (line 10).
% \vspace{-0.5em}
\vspace{-0.75em}
\begin{equation}
\begin{split}
& ran_{min} = l_i.rad - f(l_i)/2\\
& ran_{max} = l_i.rad + f(l_i)/2\\
& f(l) = (1- e^{-l.dis}) * \pi / 4.0
\end{split}
\label{range_2}
\end{equation}
\vspace{-1.5em}

We initialize the second-level sorted circle layout $MCL$ in line 12.
The second-level sorted circle layout is a concentric circle displayed in the user's view, which contains up to $N$ circles.
We initialize the current circle index $k$ of the second-level sorted circle layout $MCL$ as 0 in line 13.
Then we add all labels in the single circle layout $SCL$ to the specified circle in $MCL$ (lines 14-21).
In line 15, we use Dynamic Programming to get the longest sorted label subsequence in $SCL$. All labels in the subsequence are removed from $SCL$ and added to $MCL$.
In order to add more labels to $MCL[k]$ as {\redsout{much}}{many} as possible, we traverse each remaining label $SCL[i]$ in SCL (line 16), try to add $SCL[i]$ to $MCL[k]$ and return the updated $MCL[k]$ and $SCL$ by the function $insert$. The details of function $insert$ {\redsout{is}}{are} shown in  Algorithm \ref{alg_insert}.
After all remaining labels in SCL have performed function $insert$,
we perform $relax$ function for $MCL[k]$ to ensure that the labels arranged on $MCL[k]$ do not occlude each other (line 19). The details of $relax$ function {\redsout{is}}{are} shown in Algorithm \ref{alg_relax}.
Finally, we add $k$ to 1 (line 20). 
If there are still remaining labels in $SCL$, we will continue to iterate and try to arrange these labels in $MCL[k+1]$.

The details of the function $insert$ in Algorithm \ref{alg_layout} {\redsout{is}}{are} shown in Algorithm \ref{alg_insert}. 
The inputs of Algorithm \ref{alg_insert} are the circle layout $c$, single circle layout $SCL$, and the label $l$ need to be inserted into $c$. Algorithm \ref{alg_insert} returns the updated $c$ and the updated $SCL$.

% wxh
\vspace{-1.0em}
\begin{algorithm}
\caption{Insert}\label{alg_insert}
\KwData{circle layout $c$, $SCL$, label $l$}
\KwResult{circle layout $c$, $SCL$}
$l_l, l_r$ $\leftarrow$ binarySearch($l$, $c$)\;
% $ranI$ $\leftarrow$ $l_r.ran_{min} > l_l.ran_{max} ? [l_l.ran_{max}, l_r.ran_{min}] : \emptyset$\;
% \eIf{$l_l.rad < l_r.rad$}{\If{$l.ran \cap [l_l.rad, l_r.rad] \neq \emptyset$}{
$lInRan \leftarrow (l_l.rad \in l.ran) \lor ((l_l.rad - 2\pi) \in l.ran)$\;
$rInRan \leftarrow (l_r.rad \in l.ran) \lor ((l_r.rad - 2\pi) \in l.ran)$\;
\If{$lInRan$ $\lor$ $rInRan$}{
    $l_n \leftarrow$ nearest$(l_l, l_r, l)$\;
    % $bRad \leftarrow$ border$(l_n, l)$\;
    $l.rad \leftarrow$ median$(l_n.rad, l.ran)$\;
    % \eIf{$|l.rad - l_l.rad| < |l.rad - l_r.rad|$}
    % {$l.rad \leftarrow (l.ran_{max} + l_l.rad) /2 $\; }
    % {$l.rad \leftarrow (l.ran_{min} + l_r.rad)/2 $\;
    % }
    $c \leftarrow c + l$\;
    $SCL \leftarrow SCL - l$\;
    % \eIf{firstEle($l_r$)}{}{}
    }
return $c$, $SCL$\;
\end{algorithm}
\vspace{-1.5em}

It first uses the binary search method in alphabetical order to find the previous label $l_l$ and the next label $l_r$ in $c$ of the position where $l$ is to be inserted (line 1).
% If the radian of $l_l$ is smaller than that of $l_r$ or $l_r$ is the first element in $c$ (line 2), it tries to insert $l$ into $c$.
% It use the boolean values $lInRan$, $rInRan$ determines whether the radian of $l_l$, $l_r$ are within $l.ran$ (lines 3-4),
It {\redsout{use}}{uses} the {\redsout{boolean}}{Boolean} values $lInRan$, {and} $rInRan$ determines whether the radian of $l_l$, $l_r$ are within $l.ran$ (lines 2-3), if $lInRan$ and $rInRan$ are both true (line 5), it tries to insert $l$ into $c$.
% it calculates the current radian of $l$ for inserting $l$ into $c$ (lines 6-10). 
Specifically, it first finds the closer label $l_n$ between $l_l$ and $l_r$ according to the initial radian of $l$ (line 5), 
and then takes the median value between $l_n.rad$ and the radian of {\redsout{$l.r$ $an$}}{$l.ran$} closer to $l_n.rad$ as the final radian of $l$ (line 6).
Then it inserts $l$ into $c$ according to its updated radian (line 7), and {\redsout{remove}}{removes} $l$ from $SCL$ (line 8).
Finally, it returns the updated $c$ and the updated $SCL$ (line 10).

The details of the function $relax$ in Algorithm \ref{alg_layout} {\redsout{in}}{are} shown in Algorithm \ref{alg_relax}.
The inputs of Algorithm \ref{alg_relax} are the circle layout $c$, the circle index $k$, and the maximum number of iterations for relaxation $\#N_{it}$, {and} Algorithm \ref{alg_relax} outputs the updated $c$.

Firstly, we initialize the current number of relaxation iterations $iter$ to 0, and initialize the overlapped label array $oa$ as empty (line 1).
% 判断相邻sin(radian)相差 < 0.2 则为重叠
Then we use the function $overlappedArr$ to add all overlapped labels in $c$ into $oa$ (line 2).
In the function $overlappedArr$, we traverse all labels in $c$, add all the overlapped labels into $oa$, and sort all overlapped labels in $oa$  according to the degree of overlap from large to small (line 2).
After that, we perform the relaxation iteration until $iter \geq \#N_{it}$ or there is no overlapped label in $oa$ (lines 3-25).
In each relaxation iteration, we traverse $oa$ once (lines 4-19).
In each $oa$ traversal,
we relax the first label $oa[i]$ (lines 5-11) and the last label $oa[len(oa)-i]$ (lines 12-18) simultaneously.
For the first label $oa[i]$, we assign $oa[i]$ to $l$ (line 5).
Then we get the previous label $l_l$ of $l$ and the next label $l_r$ of $l$ in $c$ (line 6).
After that, we get the next label $l_{rr}$ of $l_r$ in $c$ (line 7).
If $l$ and $l_r$ overlap (line 8), we update the radian of $l$ by the function $subRad$ (line 9), and update the radian of $l_r$ by  the function $addRad$ (line 10). The details of the functions $subRad$ and $addRad$ {\redsout{in shown in}}{are shown in} Equation \ref{eq_subAndAddRad}.
\vspace{-0.5em}
\begin{equation}
\begin{aligned}
    subRad(l, l_l, k) = & l.rad - Min(\frac{\pi}{k*72.0},
    |l.rad - l_l.rad| - \delta,\\
    & l.rad - l.ran_{min})\\ 
    addRad(l, l_r, k) = & l.rad + Min(\frac{\pi}{k*72.0},
    |l.rad - l_r.rad| - \delta,\\ 
    & {l.ran_{max} - l.rad})
\end{aligned}
\label{eq_subAndAddRad}
\end{equation}
\vspace{-1.5em}

Then we perform the same operation on $oa[len(oa)-i]$ as $oa[i]$ (lines 12-18).
After $oa$ traversal,
we initialize $oa$ as empty, and use the function $overlappedArr$ to add all overlapped labels in $c$ into $oa$ (line 20).
If $oa$ is not empty and $iter$ is larger than $\#N_{it}$ (line 21), we remove the overlapped label $oa[0]$ with the maximum degree of overlap in $oa$ from $c$ (line 22).
Then we add $iter$ to 1 to perform the next relaxation iteration (line 24).
After all relaxation iterations are ended, we return the updated $c$ (line 26).

\vspace{-1.0em}
\begin{algorithm}
\caption{Relax}
\label{alg_relax}
\KwData{circle layout $c$, circle index $k$, maximum number of iterations for relaxation $\#N_{it}$}
\KwResult{circle layout $c$}
$iter \leftarrow 0 $; $oa$ $\leftarrow$ $\emptyset$\;
% oa: store overlap label, sorted by overlap level(compute by over label number and moving radian)
$oa \leftarrow$ overlappedArr($c$, $k$)\; 
\While{ $oa \neq \emptyset$ or iter $<$ $\#N_{it}$}{
    \For{$i \in [0,len(oa)]$}{
        $l$ $\leftarrow$ $oa[i]$\;
        $l_l$, $l_r$ $\leftarrow$ getPreNextLab($c$, $l$)\;
        $l_{rr}$ $\leftarrow$ getNextLab($c$, $l_r$)\;
        \If{overlap($l, l_r$)}{
        $l.rad \leftarrow$ subRad($l$, $l_l$, $k$)\;
        $l_r.rad \leftarrow$ addRad($l_r$, $l_{rr}$, $k$)\;
        }
        % $j \leftarrow n-i$\;
        $l$ $\leftarrow$ $oa[len(oa)-i]$\;
        $l_l$, $l_r$ $\leftarrow$ getPreNextLab($c$, $l$)\;
        $l_{rr}$ $\leftarrow$ getNextLab($c$, $l_r$)\;
        \If{overlap($l, l_l$)}{
         $l.rad \leftarrow$ subRad($l$, $l_l$, $k$)\;
        $l_r.rad \leftarrow$ addRad($l_r$, $l_{rr}$, $k$)\;
        % $l_.rad \leftarrow$ subRadian($ c[||j-1||] $)\;
        % $ c[j].rad \leftarrow$ addRadian($ c[||j+1||] $) \;
        % $ c[j].rad \leftarrow c[j].rad + $ $w(c[j])$*GetR($c[j]$, $c[j+1]$, $1$)\;
        % 加一个边界判断函数
        % $ c[j-1].rad \leftarrow c[j-1].rad - $ $w(c[j-1])$*GetR($c[j-1]$, $c[j-1-1]$, $-1$) \;
        }
    }
    % \For{$i \in [0,len(c)]$}{
    %     \If{overlap($ c[i], c[||i+1||] $)}{
    %     $c[i].rad \leftarrow$ subRadian($ c[i] $)\;
    %     $c[||i+1||].rad \leftarrow$ addRadian($ c[||i+1||] $) \;
    %     % $ c[i].rad \leftarrow c[i].rad -  w(c[i])$*GetR($c[i], c[i-1], -1$)\;
    %     % $ c[i+1].rad \leftarrow c[i+1].rad +$ $w(c[i+1])$* GetR($c[i+1+1], c[i+1], 1$) \;
    %     }
    %     $j \leftarrow n-i$\;
    %     % $mod(len(c))$;
    %     \If{overlap($ c[j], c[||j-1||]$)}{
    %     $c[||j-1||].rad \leftarrow$ subRadian($ c[||j-1||] $)\;
    %     $ c[j].rad \leftarrow$ addRadian($ c[||j+1||] $) \;
    %     % $ c[j].rad \leftarrow c[j].rad + $ $w(c[j])$*GetR($c[j]$, $c[j+1]$, $1$)\;
    %     % 加一个边界判断函数
    %     % $ c[j-1].rad \leftarrow c[j-1].rad - $ $w(c[j-1])$*GetR($c[j-1]$, $c[j-1-1]$, $-1$) \;
    %     }
    % }
    $oa$ $\leftarrow$ $\emptyset$; $oa \leftarrow$ overlappedArr($c$, $k$) \; 
    \If{$oa \neq \emptyset$ and iter $>$ $\#N_{it}$}{
        $c \leftarrow$ $c$ - $oa[0]$\;
        % \For{$i \in [0, len(oc)]$}{
        % }
    }
    $iter \leftarrow iter +1$\;
}
return $c$\;
\end{algorithm}
\vspace{-1.5em}

\subsection{View and Gaze based Label Guidance}
%  Path Generation
\label{LabelPath}
In the process of the label guidance based on the candidate labels selected in the two-level hierarchical sorted and orientated label layout, the flying trajectories, flying speed, and the locating of candidate labels {\redsout{have a great impact}}{significantly impact} on the accuracy and effectiveness of the label guidance.
For the flying trajectories of the candidate labels,
if they are too close or too far away from the user's position in VR, {\redsout{it will be difficult for the user to follow these candidate labels, which makes it difficult for the user to locate the specified object.}}{it will be challenging to follow these candidate labels, making it difficult for the user to locate the specified object.} 
For the flying speed of the candidate labels,
if they are too fast, it will be difficult for the user to keep up with {\redsout{the candidate labels}}{them}.
If the flying speed of the candidate labels {\redsout{are}}{is} too slow, {\redsout{it will reduce the efficiency of the label guidance}}{the efficiency of the label guidance will be reduced}.
For the locating of candidate labels during the label guidance,
if the user locates the wrong candidate label, {\redsout{which will result in the failure of the label guidance}}{the label guidance will fail}.

In this section, we propose the view and gaze based label guidance method for guiding the user to locate the target object in VR more efficiently.
The view and gaze based label guidance method first {generates} a specific flying trajectory for each candidate label to ensure that the distance between the candidate label and the user is even. During the process of the label guidance, the flying candidate labels will not be too close or too far away from the user.
During the process of the label guidance, this method updates the flying speed of the candidate labels to ensure that the user can keep up with the flying candidate labels efficiently.
{\redsout{And}}{Moreover,} it keeps valid candidate labels and removes the invalid {\redsout{candidate labels}}{ones} in {real time} to ensure that the user locates the correct candidate label.

% the flying speed of the candidate labels,
% And the view and gaze based label guidance method
% updates the flying speed of the candidate labels and keeps valid candidate labels in real time for guiding the user to locate the specific objects in VR more efficiently.

\begin{figure}[h]
  \vspace{-0.5em}
 \centering % avoid the use of \begin{center}...\end{center} and use \centering instead (more compact)
 \includegraphics[width=0.45\columnwidth]{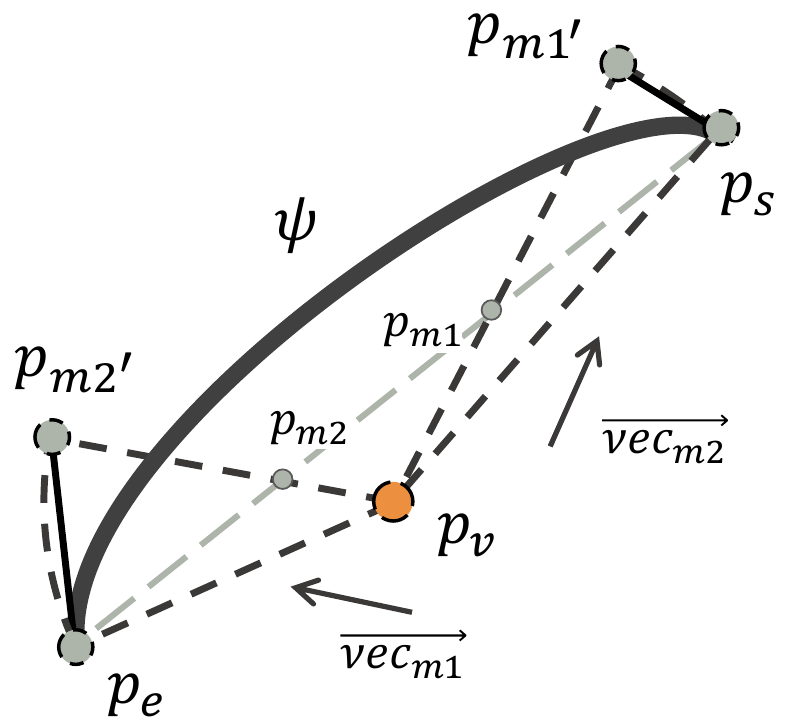}
 \caption{{\redsout{Visualization of the Flying Trajectory Generation}}{Visualization of the flying trajectory generation.}}
 \label{fig_ftg}
  \vspace{-1.5em}
\end{figure}

% wxh
We compute the specific flying trajectory $\Psi$ of each candidate label $l$ based on the initial position $p_s$, terminal position $p_e$ and the user's viewpoint position $p_v$. The details are shown in Figure \ref{fig_ftg}.
We take two trivial points $p_{m1}$ and $p_{m2}$ in the line segment formed between $p_s$ and $p_e$.
Then, we get a vector $\overrightarrow{vec_{m1}}$ passing through $p_{m1}$ with $p_v$ as the start point, the modulus length of $\overrightarrow{vec_{m1}}$ is $|p_e-p_s|$, and mark the end point of $\overrightarrow{vec_{m1}}$ as $p_{m1'}$.
Similarly, we get $\overrightarrow{vec_{m2}}$ passing through $p_{m2}$ with $p_v$ as the start point, the modulus length of $\overrightarrow{vec_{m2}}$ is $|p_v-p_e|$, and mark the end point of $\overrightarrow{vec_{m2}}$ as $p_{m2'}$.
Finally, we use $p_s$, $p_{m1'}$, $p_{m2'}$, and $p_e$ to generate a Bezier curve as $\Psi$.

During the label guidance process, we update the flying speed $s$ of the candidate label $l$.
Firstly, we get the vector that starts from the initial position $p_s$ of $l$, terminal position $p_e$ of $l$, then project the vector in the screen space to get the projected vector $\overrightarrow{vec_l}$.
Then, we record the positions of gaze point per frame, and obtain its moving direction vector $\overrightarrow{vec_g}$ fitting its position through the least square method.
We calculate the cosine value of the angle between $\overrightarrow{vec_l}$ and $\overrightarrow{vec_g}$, and get the normalized cosine value $\alpha$ by Equation \ref{equ_normCos}.
\vspace{-0.5em}
\begin{equation}
\begin{split}
\alpha & = \frac{norm(\overrightarrow{vec_l}) \cdot norm(\overrightarrow{vec_g}) +1}{2} 
\end{split}
\label{equ_normCos}
\end{equation}
\vspace{-1.5em}

The larger $\alpha$ means that the direction between $\overrightarrow{vec_l}$ and $\overrightarrow{vec_g}$ is closer, which means that the guidance effect is better and the flying speed of $l$ can be faster.
We also record the distance $dis_{lg}$ between $l$ and the gaze point $p_g$ in screen space. 
The larger $dis_{lg}$, the more difficult for the user to follow $l$. And the flying speed of $l$ should be reduced.
The flying speed $s$ of $l$ is calculated by Equation 
\ref{equ_speed}. 
\vspace{-0.5em}
\begin{equation}
\begin{split}
s & = s + (1 - s) * \alpha * (1-dis_{lg})
\end{split}
\label{equ_speed}
\end{equation}
\vspace{-1.5em}

{The symbol $s$ in {\redsout{equation}}{Equation} 4 represents the flying speed of the label, which ranges from 0 to 1. The symbol $s$ is adjusted by $1-s$, the normalized cosine value ${\alpha}$, and the distance $dis_{lg}$.
When $s$ approaches 1, we use the parameter $1-s$ for the negative feedback adjustment to slow down the increasing speed of $s$.}
{When the moving direction of the gaze point coincides more closely with the label flying trajectory, ${\alpha}$ becomes larger, and the acceleration of $s$ will be further improved. Thus we use ${\alpha}$ for the positive feedback adjustment to improve the acceleration of $s$.}
{When the distance between the label and the gaze point in screen space increases, the flying speed of the label in world space should be faster to fit the motion of the label in screen space. Thus we use the distance $dis_{lg}$ as the negative feedback adjustment to improve the acceleration of $s$.}

At last, we keep valid candidate labels and remove the invalid {\redsout{candidate labels}}{ones} during the label guidance.
For each {\redsout{fly}}{flying} candidate label $l$, if the angle between the $l$'s flying direction  $\overrightarrow{vec_{l'}}$ and the gaze moving direction $\overrightarrow{vec_{g}}$  is greater than 90$^\circ$, then $l$ has no effect on locating the target object in VR, and $l$ will be regarded as an invalid label and removed from the candidate labels in the process of label guidance.

\vspace{-0.3em}
\section{Results and Discussion: User Studies}
We design user studies to evaluate the performance of our label guidance based object locating method. 
We first design a pilot study to explore the effect of using label guidance to locate objects (Sect \ref{pilot study}). 
Then, {\redsout{we conduct a user study to further evaluate the efficiency and task load of our label guidance based method}}{we conduct a user study to further evaluate our label guidance based method's efficiency and task load}. (Sect \ref{user study}).

% We conducted the user study to evaluate the effectiveness of our two-level hierarchical sorted and oriented label layout in the object locating task. 
% We compared it to the traditional methods with no label guidance, the strictly sorted label layout and the strictly orientated label layout.

\vspace{-0.3em}
\subsection{Pilot Study}\label{pilot study}
The intuitive idea is that when there are only a few objects in the VE, the object locating task can be completed efficiently without additional guidance{\redsout{methods}} or only using a simple label guidance method.
Thus, we design a pilot study to explore the effect of using the label guidance method in the VE with {\redsout{a different number}}{different numbers} of labels.

There are two scenes and three experimental conditions in the pilot study.
The two scenes are the same except for the number of objects, 16 and 60, respectively. 
For the target object locating task, a traditional method is that users browse the entire scene to visually search for the target object (PCC1).
When there are too many objects and labels, it will be difficult to search for the target object.
A straightforward solution is to arrange the labels of objects on the screen in circular and alphabetical order, and then the user finds the target object according to the target label (PCC2).
The third condition is our label guidance based object locating method (PEC).

We recruit N=12 participants (4 female, 8 male) {\redsout{of age}}{aged} between 23 and 28. 
3 of the participants had experience with VR applications. 
We use a within-subject design with each participant completing {\redsout{each 5}}{each of the 5} tasks in 3 conditions $\times$ 2 scenes. The task is to locate the target object whose name {is} displayed at the center of the screen. Once the participant finds the target object, they will indicate it by pointing the handle ray at it and pressing a button.

\begin{table}[h]
  \vspace{-0.5em}
	\caption{Pilot study object locating time, in seconds. Statistical difference is denoted with an asterisk.}
	\label{tab:pilot}
	\scriptsize%
	\centering%
	\setlength{\tabcolsep}{2.65mm}{
	\begin{tabu}{ccccccc}
		\toprule
		\tabincell{c}{Scene} &\tabincell{c}{Condition} & \tabincell{c}{Avg \\ $\pm$ std. dev.} & \tabincell{c}{Comparison} & \tabincell{c}{$p$} \\
		
		\midrule
		\multirow{3}*{S1}
    	&$PCC1$ &	$10.5\pm6.9$ & $PCC1$ - $PEC$ & $0.025^*$  \\
		&$PCC2$ &	$9.8\pm1.8$ &  $PCC2$ - $PEC$ & $ <0.001^* $ \\
		&$PEC$&	 $6.3\pm1.2$ & &      \\
		
		\midrule
		\multirow{3}*{S2}
    	&$PCC1$ &	$22.3\pm16.3$   &$PCC1$ - $PEC$ & $<0.001^*$ \\
		&$PCC2$ &	$12.1\pm2.9$ &$PCC2$ - $PEC$ & $<0.001^*$ \\
		&$PEC$&	 $7.0\pm1.2$  & &      \\
	
		\bottomrule
		
    \end{tabu}}%
\vspace{-1.0em}
\end{table}
Tabel \ref{tab:pilot} shows the completion time for the three conditions in two scenes. The data conforms to the normal distribution by Shapiro-Wilk test\cite{1965ShaphiroTest}. 
PCC1, PCC2 are compared {\redsout{to}}{with} PEC by using ANOVA. Whether in the scene of few objects (S1) or in the scene of a large number of objects (S2), the efficiency in {\redsout{PEC}}{our method (PEC)} is significantly higher than {\redsout{PCC1 and PCC2}}{that of surrounding visually search 
 (PCC1) and arranging all labels on the screen (PCC2)}. 

The reduction in the time overhead of our method for locating the object is due to the fast search for a given label from the two-level hierarchical sorted and orientated label layout in the presence of a large number of labels.
It is necessary to apply our method.

We also compare the same method in different scenes by using ANOVA. The result is: (PCC1, $p = 0.006^*$), (PCC2, $p=0.004^*$) and (PEC, $p=0.069$). As the number of objects in the scene increases, the efficiency of traditional methods, whether {\redsout{PCC1 or PCC2}}{surrounding visually search (PCC1) or arranging all labels on the screen (PCC2)}, will become significantly worse. However, the performance of our method is independent of the number of labels.

\subsection{User study}\label{user study}

This study focuses on evaluating the efficiency and task load of our method in scenes with a large number of objects. We also evaluate the usability of our method.
We present our study using a format prescribed for health informatics evaluation reports\cite{talmon2009stare}.

\vspace{-0.3em}
\subsubsection{Study design}
In order to {\redsout{account for}}{explain} individual differences between participants, the experiment {\redsout{follow}}{follows} a full-factorial within-subject study design.
We compare our method of two-level hierarchical sorted and orientated label layout (EC3) to the conventional method of browsing the entire scene (CC1) and the straightforward method of full-screen circle layout (CC2).
Our two-level hierarchical sorted and orientated label layout has two simplified versions, one is with a single sorted circle layout in the second level (EC1), {and} the other is with a strict orientated circle layout in the second level (EC2).
% CC requires the participants not to use the label guidance method based on label layout, but to browse the entire scene to locate the object. 
In CC1, participants visually search to locate the object without the label guidance method.
CC2 places all labels on the screen, and the labels are sorted on multiple circles in alphabetical order.
EC1, EC2 and EC3 require the user to select the initial letter on the first-level sorted circle layout and then select the labels on the second-level circle layout.
The difference {\redsout{among}}{between} the three conditions lies in the second-level circle layout of labels.
{\redsout{EC1 places the labels completely in alphabetical order on a single circle without considering orientation information encoding.}}

\begin{figure}[h]
  \vspace{-0.5em}
 \centering % avoid the use of \begin{center}...\end{center} and use \centering instead (more compact)
 \includegraphics[width=0.6\columnwidth]{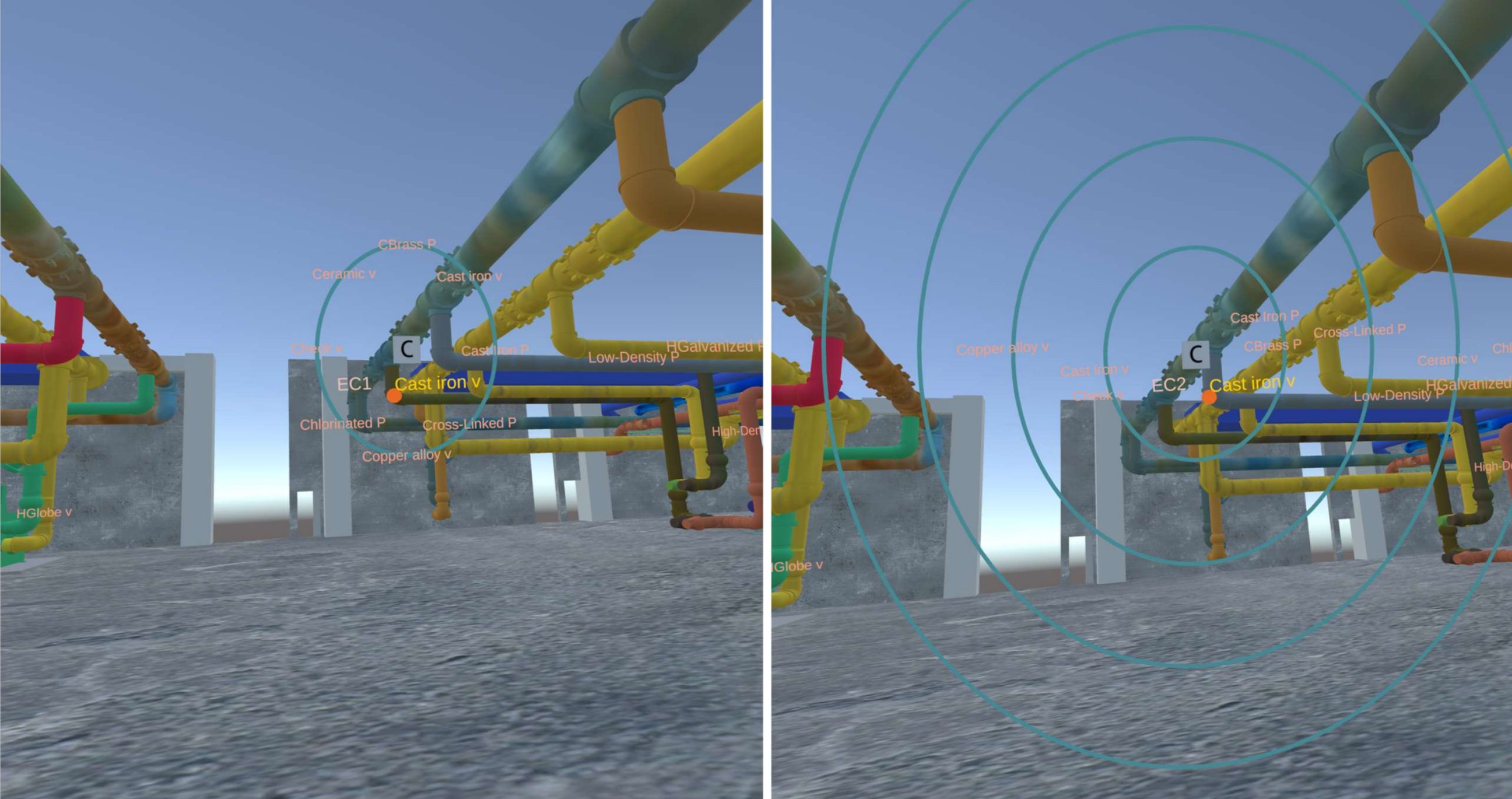}
 \caption{{EC1: single sorted circle layout in the second level(left). EC2: strict orientated circle layout in the second level (right).}}
 \label{shiyan}
  \vspace{-1.0em}
\end{figure}

{Figure \ref{shiyan} shows the visualizations of EC1 and EC2.
The idea of EC1 is consistent with that of partially-sorted concentric layout method\cite{zhou2021partially}, which places labels on a circle without overlapping according to the spring system. Compared with \cite{zhou2021partially}, EC1 also adjusts the starting position of the circle according to the distance between the labels and the anchor objects.}
EC2 places the labels on multiple circles completely according to {\redsout{the orientation information of the object}}{the object's orientation information} without considering alphabetical order information.
% 看情况，这里要不要加三个EC的图
% {\color{red} Figure\ref{xxx}}. 

\textbf{Hypotheses} 
The two-level hierarchical sorted and orientated label layout (EC3) is designed to improve the efficiency of object locating in scenes with a large number of labels. 
The two-level hierarchical layout reduces the number of labels displayed on the screen by clustering labels according to initial letter, {while} the sorted and orientated label layout reduces the layout circle number. The view and gaze based label guidance optimizes label motion trajectory and speed. Thus, we formulate the following hypotheses.

\textbf{H1}: Using the hierarchical circles to find labels (EC1, EC2, and EC3) will be \textbf{more efficient} than the traditional method (CC1 and CC2). Using the two-level hierarchical sorted and orientated label layout (EC3) will be even \textbf{more efficient} than using a single sorted circle layout (EC1) and strict orientated circle layout (EC2).

\textbf{H2}: The user \emph{task load} with the hierarchical circles (EC1, EC2, and EC3) will be lower than traditional method (CC1 and CC2), and user task load with EC3 will be lower than {those of} EC1 and EC2. 

\textbf{H3}: EC3 is \emph{easy} to use.

\begin{figure}[h]
  \vspace{-0.5em}
 \centering % avoid the use of \begin{center}...\end{center} and use \centering instead (more compact)
 \includegraphics[width=0.6\columnwidth]{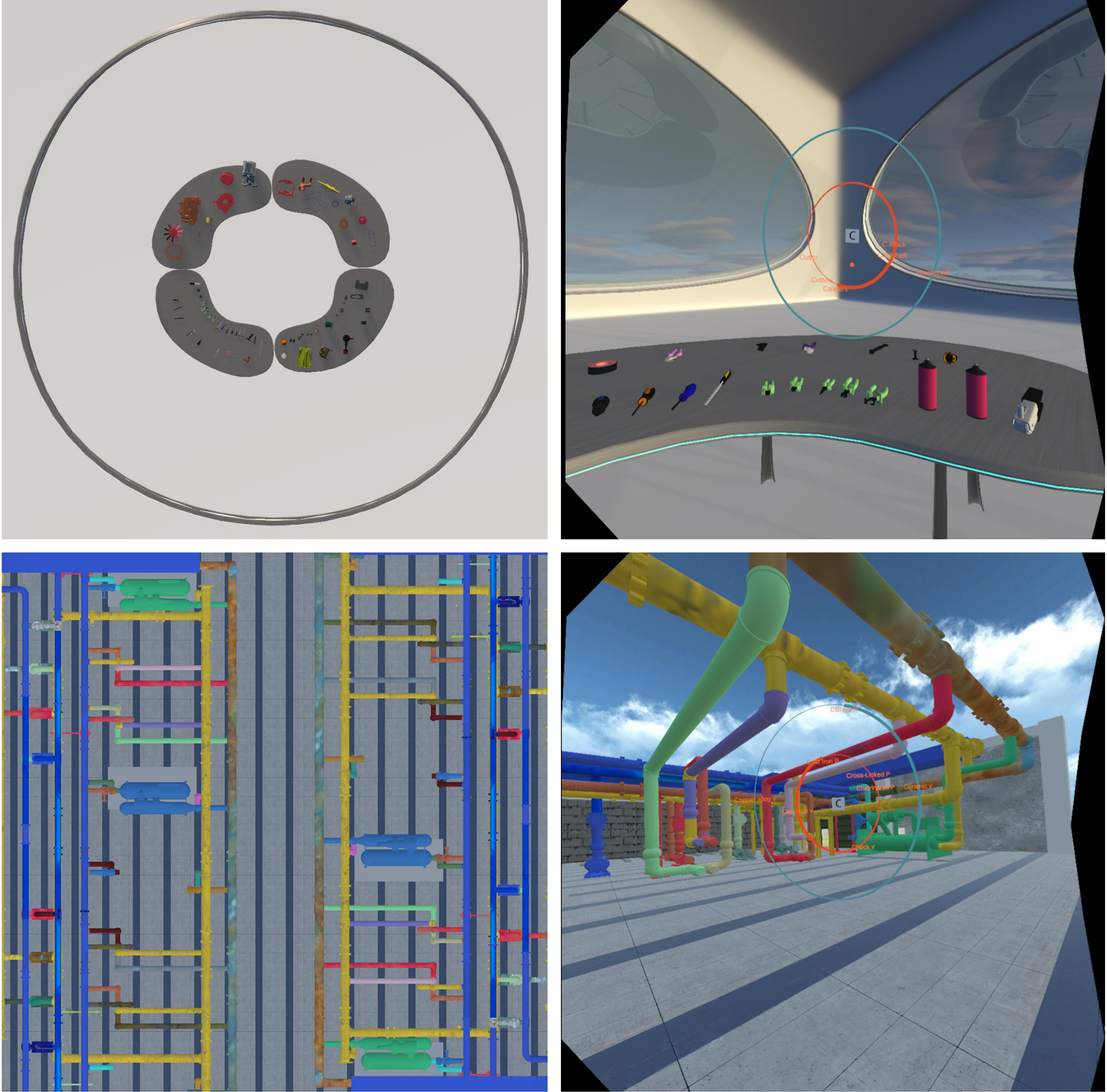}
 \caption{Two experimental scenes: workbench ({\redsout{top}}{upper} left) and pipe ({\redsout{bottom}}{lower} left). The user view of workbench ({\redsout{top}}{upper} right) and the user view of pipe ({\redsout{bottom}}{lower} right).}
 \label{scene}
  \vspace{-1.0em}
\end{figure}

\textbf{Experimental scene.} 
The experimental scenes are shown in Figure \ref{scene}.
The workbench scene is an indoor scene.
% {, and objects have roughly the same distance from the user. }
90 tools and instruments are regularly placed on the workbench. 
The pipe factory is an outdoor scene.
% and objects have different distances from the user. 
87 intricate pipes are irregularly arranged. They will partially block each other. 

{\textbf{Hardware.} We use the HTC Cosmos VR systems with two hand-held controllers, allowing the users to awake label guidance in the VE. The HMD is connected to the workstation with a 3.6GHz Intel(R) Core(TM) i7-9900KF CPU, 16GB of RAM, and an NVIDIA GeForce GTX 3080 graphics card.}

\textbf{Participants.} We recruit N=32 participants(12 female, 20 male), {\redsout{of age}}{whose ages are} between 19 and 33. 14 of our participants had experience with VR applications.
% In this experiment, due to the impact of the COVID-19, all of the participants are students and staff at our university. We took safety measures to prevent the epidemic. 

\textbf{Procedure.}
Before the experiment {\redsout{started}}{starts}, we 
% elaborate on three label layout methods, allowing 
allow participants to fully train in a simple scene until they fully understand the details of the five methods.
% 删掉的东西
% The participant presses the start key of the handle to start. The {\redsout{name of the target object}}{target object's name} will be displayed in the center of the screen. 
When the participant {\redsout{found}}{finds} the target object, he should point the handle ray at the target object and presses the end key to complete an object locating trial and then a prompt to continue the next trial {\redsout{would}}{will} appear in the center of the screen. The participant repeats the above steps. 
During the whole process, the participant can only turns around and moves his head.
We balance the order of each condition in a Latin square ({\redsout{4 groups}}{5 groups}). Each participant completes 150 study trials: 5 conditions $\times$ 2 scenes $\times$ 15 trials. The participant can take breaks between each trial. After the experiment, The participant fills out subjective post-experiment questionnaires.

% Before , We introduced the experiment scene and process in detail.
% The experimental procedure starts with 
% The task was to locate 15 objects

% A task flow was: We randomly displayed a string on the screen for the participant to find the corresponding object. Participants pressed the start key on handle to start CC trial or to call a label layout methods to start EC. The conditions of EC was controlled by program. Participants should search and track the corresponding label as quickly as possible. Finally, participants pointed the handle ray at the anchor object as precisely to confirm his result.

% They could take breaks between each task. After the experiment, participants filled out subjective post-experiment questionnaires.

\textbf{Metrics.} 
We use two objective metrics: $time$ and the $rotation\ angle$. 
$time$ to locate an object in each trial starts when the participant presses the start key of the handle and ends when the participant points to the target object and presses the end key. 
We record the cumulative $fov Time$ of the target object in the central FOV to remove outliers. If the $fov Time$ takes up more than 50\% part, we abandon the data of this trial. 
We use $rotation\ angle$ to record the accumulated head rotation angle of the participant. The moment to record $rotation\ angle$ is {\redsout{as}}{the} same as recording $time$. 

We record the {\redsout{perception}}{perceived} value with two subjective metrics: user task load, measured with a standard NASA TLX questionnaire \cite{Hart2006NasaTaskLI_r50, Hart1988DevelopmentON_r51} and usability of our method, measured with a 5-point Likert scale question.
The Likert scale has five questions. Each question needs users to score from 1 to 5. 1 means extremely disagree, and 5 means extremely agree. These five {\redsout{problems}}{questions} are: Q1: {whether} the two-level hierarchical sorted and orientated label layout(EC3) can effectively improve the \emph{efficiency}? Q2: {whether} EC3 is very simple and \emph{easy} to learn? Q3: {whether} EC3 dose not \emph{confuse} you when completing tasks? Q4: {whether} EC3 is convincing and \emph{reasonable}? Q5: {whether} you {\redsout{are \emph{enjoyed} to use}}{\emph{enjoy} using} EC3 in the future?

\textbf{Statistical analysis.} The time to locate the target object, the task load and the usability scores are compared across the five conditions (CC1, CC2, EC1, EC2, and EC3) by using a one-way repeated measures ANOVA. Firstly, we use the Shapiro-Wilk test\cite{1965ShaphiroTest} to verify the distribution normality assumption. Secondly, we use the Mauchly test\cite{mauchly1940significance} to verify the sphericity assumption. When the sphericity assumption is violated, we apply a Greenhouse-Geisser correction.
Then, {\redsout{We conducted}}{we conduct} a population ANOVA to investigate whether the null hypothesis of no statistically significant difference between the five conditions could be rejected.
When null hypothesis {\redsout{was}}{is} rejected ($p < 0.05$), {the} differences between the four pairs (EC3 vs CC1, CC2, EC1, EC2) are analyzed by post-hoc tests, using the Bonferroni correction to reduce the level of significance ($p < 0.016$).
We use Cohen's d \cite{JCohen2013Cohensd_r48} to quantify the effect size.

\vspace{-0.3em}
\subsubsection{ Experimental results }
The time and the pairwise comparisons among conditions are displayed in Tabel \ref{tab:time}. 
The time in {the} workbench and pipe both violate the sphericity assumption. After applying the Greenhouse-Geisser correction, the overall ANOVA reveals significant differences between the five conditions:($F_{1.126, 27.036} = 27.768$, $p<0.001$) for the workbench scene and ($F_{1.420, 73.860} = 56.009$, $p<0.001$) for the pipe scene. 
Post-hoc analysis shows that {\redsout{EC3 time}}{the time of EC3} is significantly shorter than {that of} CC1, CC2, EC1, and EC2 in both scenes. The effect size for all pairwise is V.large or higher.

\begin{table}[h]
  \vspace{-0.5em}
	\caption{The time to locate object, in seconds. Statistical significance ($p < 0.016$) is denoted with an asterisk.}
	\label{tab:time}
	\scriptsize%
	\centering%
	\setlength{\tabcolsep}{1.2mm}{
	\begin{tabu}{ccccccc}
		\toprule
		\tabincell{c}{Scene} & \tabincell{c}{Condi\\-tion} & \tabincell{c}{Avg \\ $\pm$ std. dev.}
		& \tabincell{c}{($XC_i$-$EC_3$)\\/ $XC_i$} & \tabincell{c}{$p$} &\tabincell{c}{Cohen's \\ $d$} & \tabincell{c}{Effect \\ size} \\
		
		\midrule
		\multirow{5}*{S1}
    	&$CC1$ &	$25.4\pm12.9$   &   74.2$\%$
		& $<0.001^*$  &  $2.05$ & Huge \\
		&$CC2$ &	$11.6\pm3.89$   &   43.7$\%$
		& $<0.001^*$  &  $1.75$ & V.large \\
		&$EC1$ &	$8.8\pm1.6$   &   25.2$\%$
		& $ <0.001^* $  &  $1.51$ & V.large \\
		&$EC2$ &	$8.4\pm1.4$   &   21.6$\%$
		& $<0.001*$  &  $1.31$ & V.large \\
		&$EC3$&	 $6.6\pm1.3$  & & & &      \\
		
		\midrule
		\multirow{5}*{S2}
    	&$CC1$ &	$19.9\pm8.5$   &   64.1$\%$
		& $<0.001^*$  &  $2.07$ & Huge \\
		&$CC2$ &	$13.4 \pm 2.9$   &   $46.6\%$
		& $<0.001^*$  &  $2.47$ & Huge \\
		&$EC1$ &	$10.9\pm2.2$   &   34.6$\%$
		& $<0.001^*$  &  $1.76$ & V.large \\
		&$EC2$ &	$10.3\pm2.4$   &   30.8$\%$
		& $<0.001^*$  &  $1.39$ & V.large \\
		&$EC3$&	 $7.1\pm2.1$  & & & &      \\
	
		\bottomrule
		
    \end{tabu}}%
	
\vspace{-2.0em}
\end{table}

\begin{table}[h]
	\caption{The Rotation angle during object locating, in degrees. Statistical significance ($p < 0.016$) is denoted with an asterisk.}
	\label{tab:rotation}
	\scriptsize%
	\centering%
	\setlength{\tabcolsep}{1.2mm}{
	\begin{tabu}{ccccccc}
		\toprule
		\tabincell{c}{Scene} & \tabincell{c}{Condi\\-tion} & \tabincell{c}{Avg \\ $\pm$ std. dev.}
		& \tabincell{c}{($XC_i$-$EC_3$)\\/ $XC_i$} & \tabincell{c}{$p$} &\tabincell{c}{Cohen's \\ $d$} & \tabincell{c}{Effect \\ size} \\
		
		\midrule
		\multirow{5}*{S1}
    	&$CC1$ &	$642\pm379$   &   87.9$\%$
		& $<0.001^*$  &  $2.09$ & Huge \\
		&$CC2$ &	$114.9\pm45.5$   &   32.5$\%$
		& $0.010^*$  &  $1.04$ & Medium \\
		&$EC1$ &	$99.4\pm37.7$   &   21.9$\%$
		& $ 0.011* $  &  $0.64$ & Medium \\
		&$EC2$ &	$75.2\pm29.4$   &   -3.32$\%$
		& $0.509$  &  $0.08$ & V.small \\
		&$EC3$&	 $77.6\pm29.6$  & & & &      \\
		
		\midrule
		\multirow{5}*{S2}
    	&$CC1$ &	$448\pm494$   &   81.9$\%$
		& $<0.001*$  &  $1.05$ & Large \\
		&$CC2$ &	$112.8\pm30.5$   & 27.8$\%$
		& $<0.001^*$  &  $1.53$ & V.large \\
		&$EC1$ &	$99.9\pm32.7$   &   18.6$\%$
		& $<0.001^*$  &  $0.55$ & Medium \\
		&$EC2$ &	$78.7\pm36.2$   &   -3.43$\%$
		& $0.519$  &  $0.077$ & V.small \\
		&$EC3$&	 $81.4\pm34.0$  & & & &      \\
	
		\bottomrule
		
    \end{tabu}}%
	
  % \vspace{-0.0em}
\end{table}
\vspace{-1.0em}

The rotation angle and the pairwise comparisons among conditions are displayed in Tabel \ref{tab:rotation}. 
The rotation angle in workbench scene and pipe scene both violate the sphericity assumption. After applying the Greenhouse-Geisser correction, the overall ANOVA reveals significant differences between the five conditions:($F_{1.029, 51.456} = 100.676$, $p<0.001$) for the workbench scene and ($F_{1.028, 44.220} = 22.583$, $p<0.001$) for the pipe scene. 
Post-hoc analysis shows that EC3 rotation angle is significantly smaller than CC1, CC2 and EC1 but not significantly different from EC2. The effect size between EC3 and CC1 is "Large" or higher. The effect size between EC3 and CC2 is "Very large" for pipe scene. 
% The rotation angle in EC was significantly smaller than those in CC. The angle data in EC3 was significantly smaller than those in EC1, but the effect size in pairwise comparisons between EC3 and EC1 was "Medium". The angle data in EC3 was no significantly different with those in EC2.

\begin{figure}[h]
  \vspace{-0.5em}
 \centering % avoid the use of \begin{center}...\end{center} and use \centering instead (more compact)
 \includegraphics[width=0.95\columnwidth]{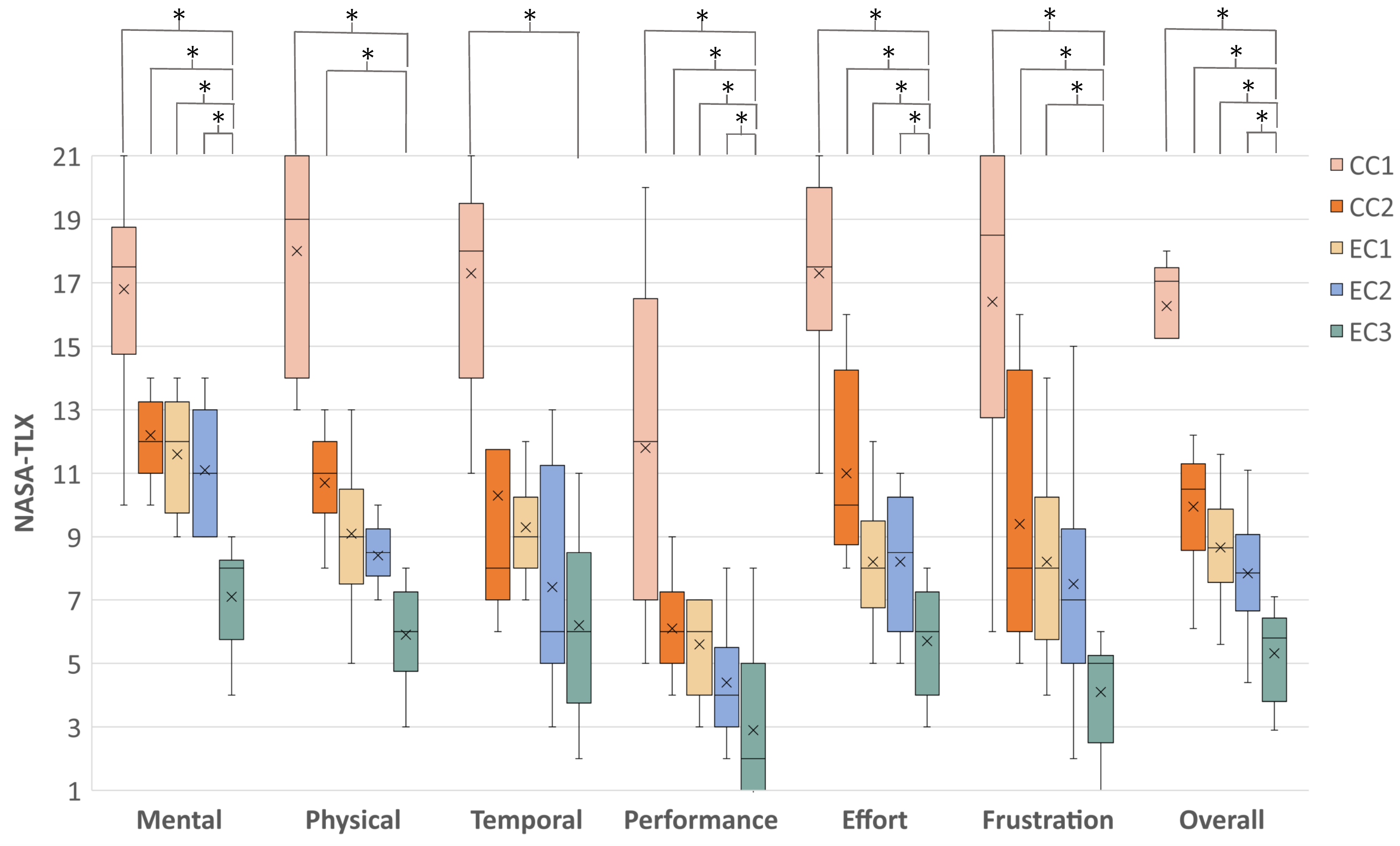}
 \caption{NASA-TLX scores for individual questions. Significant difference are denoted with the asterisk and line.}
 \label{fig:nasa}
  \vspace{-1.0em}
\end{figure}

The NASA-TLX scores is shown on Figure \ref{fig:nasa}. The positive score \emph{{\redsout{Performance}}{performance}} is replaced with its complement {\redsout{such}}{so} that smaller is always more favorable. We {\redsout{tested}}{test} the sphericity assumption on six aspects of NASA-TLX and {\redsout{applied}}{apply} the Greenhouse-Geisser correction when necessary. The \emph{mental} is verified with $p=0.666$, the \emph{physical} is verified with $p=0.546$, the \emph{temporal} is verified with $p=0.972$, the \emph{performance} is violated with $p<0.001$, the \emph{effort} is verified with $p=0.327$ and the \emph{frustration} is verified with $p=0.211$.
After applying the Greenhouse-Geisser correction on \emph{performance}, the overall ANOVA reveals significant differences with $(F_{1.097, 5.485}=9.516,p=0.023)$. 
Compared with CC1, EC3 has significant improvement in all six aspects. Compared with CC2, EC3 has significant improvement in five aspects except {in} temporal. Compared with EC1, EC3 {\redsout{has significant improvement in mental}}{significantly improves mental,} performance, effort, and frustration. Compared with EC2, EC3 has significant improvement in mental, performance and effort. For the overall score, EC3 has significant improvement compared with CC1, CC2, EC1, and EC2.

% \begin{figure}[h]
%  \centering % avoid the use of \begin{center}...\end{center} and use \centering instead (more compact)
%  \includegraphics[width=\columnwidth]{2.pdf}
%  \caption{the NASA-TLX result}
%  \label{fig:nasa}
% \end{figure}

We design a Likert scale to investigate the usability of our two-level hierarchical sorted and orientated label layout. The results show that participants believe that EC3 {\redsout{could}}{can} help them improve \emph{efficiency} (4.92), {and} they feel \emph{easy} (4.25) and comfortable when using EC3. Overall, EC3 does \emph{not confuse} them (3.89). The layout of EC3 is \emph{reasonable} (4.25), and they \emph{enjoy} (4.88) {\redsout{to use}}{using} it in the future.
% We designed a Likert scale to investigate users' subjective evaluation of two-level hierarchical sorted and orientated label layou(EC3). This scale has five questions. Each question needs users to score from 1 to 5. 1 means extremely disagree and 5 means extremely agree. These five problems are: Q1: locating objects through two-level hierarchical sorted and orientated label layout can effectively improve the efficiency; Q2: two-level hierarchical sorted and orientated label layout is very simple and easy to learn; Q3: two-level hierarchical sorted and orientated label layout dose not confuse you when completing tasks; Q4: two-level hierarchical sorted and orientated label layout is convincing and reasonable; Q5: you are willing to use two-level hierarchical sorted and orientated label layout in the future.

\vspace{-0.3em}
\subsubsection{Discussion}
% 按照假设的结论组织。每个结论一段话，如何根据数据得出来的
% 再重新组织一下
% {\color{red}Discussion}
The study is conducted to evaluate the performance of our two-level hierarchical sorted and orientated label layout for the object locating task in VR. 
The experimental results show that our method achieves a significant improvement over {\redsout{the}} other methods in both efficiency and task load, and our method is easy to use.

% H1 改成more efficient
% 时间和角度
% T2T3放在一起说，
% Ec2Ec3rotation，造成原因；虽然转角更小，没有显著差别；因为这个原因H1不成立
% 先说AA，再说速度
% H1不成立，P值相差很小
% 所有假设后面说总结

For \textbf{H1},
{the} mean value and standard deviation in Table \ref{tab:time} {\redsout{shows}}{show} that using the hierarchical circles (EC1, EC2 and EC3) is much faster and requires {\redsout{less}}{fewer} head rotation angles than the traditional methods (CC1 and CC2). 
ANOVA analysis shows that EC3 is significantly faster than other methods and significantly reduces the accumulated head rotation angle compared with CC1, CC2 and EC1. 
However, there is no significant difference in the rotation angle between EC2 and EC3. 
EC3 is \emph{more efficient} than  CC1, CC2, EC1, but not \emph{more efficient} than EC2 because of no significant difference in the rotation angle.
The results do not support \textbf{H1}.

% The result do not support \emph{H1}, with significantly more efficient for EC3 vs CC1, CC2, EC1, but not more efficient for EC3 vs EC2.
% The mean value and standard deviation in Table \ref{tab:time} shows that using the hierarchical circles (EC1, EC2 and EC3) to locate objects is much faster and requires less head rotation angles than the traditional methods (CC1 and CC2). 
% The ANOVA analysis shows that EC3 is significantly faster than other methods and significantly reduces the accumulated head rotation angle compared with CC1, CC2 and EC1. 
% However, there is no significant difference in the rotation angle between EC2 and EC3 and EC3 is worse than EC2. Therefore EC3 is not \emph{more efficient} than EC2. \emph{H1} is not supported.
Firstly, we discuss $time$.
Compared with {the} traditional {\redsout{method}}{methods}, the user only needs to retrieve the initial clustering labels to find the target label instead of changing his perspective to retrieve all labels.
Compared with EC1, EC3 encodes the orientation information, and thus EC3 places the labels closer to corresponding objects.
% adaptive label range controlled circle layout 
Compared with EC2, the insertion and relaxation in EC3 reduce the number of label layout circles. 
Compared with one label selection based on dwell-time in EC1 and EC2, EC3 selects multiple candidate labels based on {the} movement of gaze point.
To sum up, EC3 is significantly faster than other {\redsout{method}}{methods}.
% In addition, two-level hierarchical sorted and orientated label layout (EC3) can use the moving direction of gaze point for candidate label selection and label tracking operations, while single sorted circle label layout (EC1) and strict orientated label layout (EC2) only select one label via the dwell time method. This step also speeds up object locating.
Secondly, we discuss $rotation\ angle$. CC1, CC2 and EC1 do not consider the orientation between the label and the object. The two processes of the user searching the label and locating the object are {\redsout{not related}}{unrelated}. In EC2 and EC3, the orientation of label and the gaze point is basically the same as object and the gaze point, so no additional head rotation angle will be introduced. However, compared with the strict orientated circle layout in the second level (EC2), changing the position of label on the circle in EC3 may cause the guidance path to be not optimal, which introduces {an} additional rotation angle.

For \textbf{H2},
Figure \ref{fig:nasa} shows that the mean score of EC1, EC2 and EC3 in {\redsout{of}} six aspects is lower than that of CC1 and CC2. Compared with other methods, EC3 has significantly improvement in \emph{mental}, \emph{performance} and \emph{frustration}. 
The {\redsout{EC3 task}} load of EC3 is \emph{significantly lower} than that of  CC1, CC2, EC1, and EC2.
The results support \textbf{H2}.

% The results support \emph{H2}, with significantly lower task load for EC3 vs CC1, CC2, EC1 and EC2.
% The Figure \ref{fig:nasa} shows that the mean score of EC1, EC2 and EC3 in all of six aspects is lower than CC1 and CC2. And EC3 has significantly improvement in \emph{mental}, \emph{performance} and \emph{frustration} compared with other method. 

CC1 requires the user to wrap around and change his perspective to search, memorize and compare from different perspectives, {and} thus the task load of CC1 is {the} highest.
CC2 arranges all labels on one screen. Limited by the FOV of the device, the user has to change his perspectives to find the target label. In addition, the selection method based {\redsout{dwell-time}}{on dwell time} also brings trouble to users, which {\redsout{makes users have to}}{makes it necessary for users to} move the gaze point carefully to select.
The label layout of EC3 is more reasonable and the label selection of EC3 is more natural, thus the task load of EC3 in \emph{mental}, \emph{effort} and \emph{frustration} is significantly lower than {that of} EC1 and EC2. 
The discussion about \textbf{H1} can explain that the task load of EC3 in \emph{performance} is significantly lower than EC1 and EC2. 
% Compared with EC2, EC3 has fewer circles, and the selection strategy of EC3 is more natural and comfortable, which can reduce the burden on users.

For \textbf{H3}, the \emph{effort} of NASA-TLX task load shows that the method in EC3 requires less effort.
The results support \textbf{H3}.
The result of Likert scale questions shows that users believe that EC3 is easy to learn and {\redsout{they enjoy to use}}{that they will enjoy using} EC3 in the future.

% The results support \emph{H3}, the \emph{effort} of NASA-TLX task load shows that the method in EC3 requires less effort. The result of Likert scale questions shows that users believe that EC3 is easy to learn and they enjoy to use EC3 in the future.

% In this section, we propose the view and gaze based label guidance method for guiding the user to locate the target object in VR more efficiently.
% The view and gaze based label guidance method  first generates a specific flying trajectory for each candidate label to ensure that the distance between the candidate label and the user is even and during the process of the label guidance, and the flying candidate labels will not be too close or too far away from the user.
% During the process of  the label guidance, this method updates the flying speed of the candidate labels to ensure that the user can keep up with the flying candidate labels efficiently.
% And it keeps valid candidate labels and drop the invalid candidate labels in real time to ensure that the user locates the correct candidate label.

\vspace{-0.3em}
\section{Conclusion, limitations and future work}
% the label guidance based object locating method
% the two-level hierarchical sorted and orientated label layout
% the view and gaze based label guidance method

 We have proposed an efficient object locating method based on label guidance to improve the efficiency of locating the target object in VR applications.
 A two-level hierarchical sorted and orientated label layout is designed to provide sort and orientation cues of the target object.
 A view and gaze label guidance method is introduced to improve the efficiency of locating {the} target object.
 Our method achieves a significant improvement in efficiency and a significant reduction of task load.

%  - Our method will be less efficient when many labels have the same initial.
%  二级标签layout复杂，圈数多，即使排序仍然难找，task load，找标签依然花费很多时间，效率降低

% - Our method may cause confuse when multiple objects locating together and with similar labels.
% 如果两个标签名字很近，物体靠在一起，飞行路径非常相似，有效标签也不止一个，用户找起来有困惑
% - when multiple similar path, introduce a diverse path to remove other possibility.

% - we didn't solve the complete occlusion problem 
% object and label are behind another object.
% 没有处理完全遮挡问题，物体和标签被另一个物体完全挡住
% - Occlusion  label always in view.

% - future work , the current method has two step for object locating: find label and locate the object with label guidance,
% one step
% % 把外标签转成内标签 embeded label
% % 建立一个连线，把所有内标签连线

There are some limitations {\redsout{to}}{in} our method.
The first limitation is {that} the efficiency of our method will reduce in the VE that has many labels with the same initials.
Although all labels are arranged in alphabetical order on each circle of the second-level sorted circle layout, {\redsout{but}} the second-level sorted circle layout will contain many circles in this case, so it still takes {\redsout{many}}{much} time for the user to find the candidate labels.
Thus, one possible future work is to propose an adaptive range calculation method, which dynamically expands the range of the labels {\redsout{that have the same initials so as}}{with the same initials}.
% 删掉的东西
% {to reduce the number of circles of the second-level sorted circle layout.}
The second limitation is that the flying trajectories of some candidate labels may {be} similar
when there are many objects with the same initials in the same region of the VE, which will reduce the accuracy of the object locating.
The second future work is to introduce a trajectory similarity parameter.
When there are multiple similar flying trajectories, these trajectories are deformed by the trajectory similarity parameter to ensure the uniqueness of each trajectory.
The third limitation is that our method can not guide the user to locate the target object efficiently when {\redsout{the target is completely occluded by the other object}}{the other object completely occludes the target}.
So another possible future work is to combine our method with multiperspective visualization to remove occlusions of the target object in the process of label guidance.
{The fourth limitation is the relationship between label placement and label size. We slide the handle button to resize labels so that the labels can display information clearly without overly occluding the scene. We fix the size of the labels before the user study. We do not consider the dynamic size, which needs to be synchronized and enlarged when moving to the target object.}

\vspace{-0.3em}
\acknowledgments{
This work was supported by National Key R\&D plan 2019YFC1521102, by the National Natural Science Foundation of China through Projects 61932003 and 61772051.}

% %% if specified like this the section will be committed in review mode
% \acknowledgments{
% The authors wish to thank A, B, and C. This work was supported in part by
% a grant from XYZ.}

%\bibliographystyle{abbrv}
\bibliographystyle{abbrv-doi}

\bibliography{template}

\begin{thebibliography}{10}

\bibitem{bodonyi2020efficient}
A.~Bodonyi and R.~Kunkli.
\newblock Efficient object location determination and error analysis based on
  barycentric coordinates.
\newblock {\em Visual Computing for Industry, Biomedicine, and Art}, 3(1):1--7,
  2020.

\bibitem{bork2018towards}
F.~Bork, C.~Schnelzer, U.~Eck, and N.~Navab.
\newblock Towards efficient visual guidance in limited field-of-view
  head-mounted displays.
\newblock {\em IEEE transactions on visualization and computer graphics},
  24(11):2983--2992, 2018.

\bibitem{cmolik2020mixed}
L.~Cmolik, V.~Pavlovec, H.-Y. Wu, and M.~Nollenburg.
\newblock Mixed labeling: Integrating internal and external labels.
\newblock {\em IEEE Transactions on Visualization and Computer Graphics}, 2020.

\bibitem{JCohen2013Cohensd_r48}
J.~Cohen.
\newblock Statistical power analysis for the behavioral sciences.
\newblock Academic press, 2013.

\bibitem{fink2012algorithms}
M.~Fink, J.-H. Haunert, A.~Schulz, J.~Spoerhase, and A.~Wolff.
\newblock Algorithms for labeling focus regions.
\newblock {\em IEEE Transactions on Visualization and Computer Graphics},
  18(12):2583--2592, 2012.

\bibitem{grasset2012image}
R.~Grasset, T.~Langlotz, D.~Kalkofen, M.~Tatzgern, and D.~Schmalstieg.
\newblock Image-driven view management for augmented reality browsers.
\newblock In {\em 2012 IEEE International Symposium on Mixed and Augmented
  Reality (ISMAR)}, pp. 177--186. IEEE, 2012.

\bibitem{gruenefeld2017eyesee360}
U.~Gruenefeld, D.~Ennenga, A.~E. Ali, W.~Heuten, and S.~Boll.
\newblock Eyesee360: Designing a visualization technique for out-of-view
  objects in head-mounted augmented reality.
\newblock In {\em Proceedings of the 5th symposium on spatial user
  interaction}, pp. 109--118, 2017.

\bibitem{gruenefeld2018flyingarrow}
U.~Gruenefeld, D.~Lange, L.~Hammer, S.~Boll, and W.~Heuten.
\newblock Flyingarrow: pointing towards out-of-view objects on augmented
  reality devices.
\newblock In {\em Proceedings of the 7th ACM international symposium on
  pervasive displays}, pp. 1--6, 2018.

\bibitem{Hart2006NasaTaskLI_r50}
S.~Hart.
\newblock Nasa-task load index (nasa-tlx); 20 years later.
\newblock {\em Proceedings of the Human Factors and Ergonomics Society Annual
  Meeting}, 50:904 -- 908, 2006.

\bibitem{Hart1988DevelopmentON_r51}
S.~Hart and L.~Staveland.
\newblock Development of nasa-tlx (task load index): Results of empirical and
  theoretical research.
\newblock {\em Advances in psychology}, 52:139--183, 1988.

\bibitem{heinsohn2014boundary}
N.~Heinsohn, A.~Gerasch, and M.~Kaufmann.
\newblock Boundary labeling methods for dynamic focus regions.
\newblock In {\em 2014 IEEE Pacific Visualization Symposium}, pp. 243--247.
  IEEE, 2014.

\bibitem{jia2021semantic}
J.~Jia, S.~Elezovikj, H.~Fan, S.~Yang, J.~Liu, W.~Guo, C.~C. Tan, and H.~Ling.
\newblock Semantic-aware label placement for augmented reality in street view.
\newblock {\em The Visual Computer}, 37(7):1805--1819, 2021.

\bibitem{kouvril2018labels}
D.~Kou{\v{r}}il, L.~{\v{C}}mol{\'\i}k, B.~Kozl{\'\i}kov{\'a}, H.-Y. Wu,
  G.~Johnson, D.~S. Goodsell, A.~Olson, M.~E. Gr{\"o}ller, and I.~Viola.
\newblock Labels on levels: labeling of multi-scale multi-instance and crowded
  3d biological environments.
\newblock {\em IEEE transactions on visualization and computer graphics},
  25(1):977--986, 2018.

\bibitem{kruijff2018influence}
E.~Kruijff, J.~Orlosky, N.~Kishishita, C.~Trepkowski, and K.~Kiyokawa.
\newblock The influence of label design on search performance and noticeability
  in wide field of view augmented reality displays.
\newblock {\em IEEE transactions on visualization and computer graphics},
  25(9):2821--2837, 2018.

\bibitem{lehtinen2012dynamic}
V.~Lehtinen, A.~Oulasvirta, A.~Salovaara, and P.~Nurmi.
\newblock Dynamic tactile guidance for visual search tasks.
\newblock In {\em Proceedings of the 25th annual ACM symposium on User
  interface software and technology}, pp. 445--452, 2012.

\bibitem{lin2021labeling}
T.~Lin, Y.~Yang, J.~Beyer, and H.~Pfister.
\newblock Labeling out-of-view objects in immersive analytics to support
  situated visual searching.
\newblock {\em IEEE Transactions on Visualization and Computer Graphics}, 2021.

\bibitem{lindeman2003effective}
R.~W. Lindeman, Y.~Yanagida, J.~L. Sibert, and R.~Lavine.
\newblock Effective vibrotactile cueing in a visual search task.
\newblock In {\em Proc. of Interact 2003}, pp. 89--96, 2003.

\bibitem{marquardt2020comparing}
A.~Marquardt, C.~Trepkowski, T.~D. Eibich, J.~Maiero, E.~Kruijff, and
  J.~Sch{\"o}ning.
\newblock Comparing non-visual and visual guidance methods for narrow field of
  view augmented reality displays.
\newblock {\em IEEE Transactions on Visualization and Computer Graphics},
  26(12):3389--3401, 2020.

\bibitem{mauchly1940significance}
J.~W. Mauchly.
\newblock Significance test for sphericity of a normal n-variate distribution.
\newblock {\em The Annals of Mathematical Statistics}, 11(2):204--209, 1940.

\bibitem{mcintire2010visual}
J.~P. McIntire, P.~R. Havig, S.~N. Watamaniuk, and R.~H. Gilkey.
\newblock Visual search performance with 3-d auditory cues: Effects of motion,
  target location, and practice.
\newblock {\em Human factors}, 52(1):41--53, 2010.

\bibitem{mcnamara2019information}
A.~McNamara, K.~Boyd, J.~George, W.~Jones, S.~Oh, and A.~Suther.
\newblock Information placement in virtual reality.
\newblock In {\em 2019 IEEE Conference on Virtual Reality and 3D User
  Interfaces (VR)}, pp. 1765--1769. IEEE, 2019.

\bibitem{peterson2008managing}
S.~Peterson, M.~Axholt, and S.~R. Ellis.
\newblock Managing visual clutter: A generalized technique for label
  segregation using stereoscopic disparity.
\newblock In {\em 2008 IEEE Virtual Reality Conference}, pp. 169--176. IEEE,
  2008.

\bibitem{renner2017attention}
P.~Renner and T.~Pfeiffer.
\newblock Attention guiding techniques using peripheral vision and eye tracking
  for feedback in augmented-reality-based assistance systems.
\newblock In {\em 2017 IEEE symposium on 3D user interfaces (3DUI)}, pp.
  186--194. IEEE, 2017.

\bibitem{schwerdtfeger2008supporting}
B.~Schwerdtfeger and G.~Klinker.
\newblock Supporting order picking with augmented reality.
\newblock In {\em 2008 7th IEEE/ACM International Symposium on Mixed and
  Augmented Reality}, pp. 91--94. IEEE, 2008.

\bibitem{1965ShaphiroTest}
S.~S. Shaphiro and M.~B. Wilk.
\newblock An analysis of variance test for normality (complete samples).
\newblock {\em Biometrika}, 52:591--611, 1965.

\bibitem{talmon2009stare}
J.~Talmon, E.~Ammenwerth, J.~Brender, N.~De~Keizer, P.~Nyk{\"a}nen, and
  M.~Rigby.
\newblock Stare-hi—statement on reporting of evaluation studies in health
  informatics.
\newblock {\em International journal of medical informatics}, 78(1):1--9, 2009.

\bibitem{tatzgern2014hedgehog}
M.~Tatzgern, D.~Kalkofen, R.~Grasset, and D.~Schmalstieg.
\newblock Hedgehog labeling: View management techniques for external labels in
  3d space.
\newblock In {\em 2014 IEEE Virtual Reality (VR)}, pp. 27--32. IEEE, 2014.

\bibitem{tatzgern2013dynamic}
M.~Tatzgern, D.~Kalkofen, and D.~Schmalstieg.
\newblock Dynamic compact visualizations for augmented reality.
\newblock In {\em 2013 IEEE Virtual Reality (VR)}, pp. 3--6. IEEE, 2013.

\bibitem{tatzgern2016adaptive}
M.~Tatzgern, V.~Orso, D.~Kalkofen, G.~Jacucci, L.~Gamberini, and
  D.~Schmalstieg.
\newblock Adaptive information density for augmented reality displays.
\newblock In {\em 2016 IEEE Virtual Reality (VR)}, pp. 83--92. IEEE, 2016.

\bibitem{trepkowski2019effect}
C.~Trepkowski, D.~Eibich, J.~Maiero, A.~Marquardt, E.~Kruijff, and S.~Feiner.
\newblock The effect of narrow field of view and information density on visual
  search performance in augmented reality.
\newblock In {\em 2019 IEEE Conference on Virtual Reality and 3D User
  Interfaces (VR)}, pp. 575--584. IEEE, 2019.

\bibitem{van2008pip}
E.~Van~der Burg, C.~N. Olivers, A.~W. Bronkhorst, and J.~Theeuwes.
\newblock Pip and pop: nonspatial auditory signals improve spatial visual
  search.
\newblock {\em Journal of Experimental Psychology: Human Perception and
  Performance}, 34(5):1053, 2008.

\bibitem{xu2019ringtext}
W.~Xu, H.-N. Liang, Y.~Zhao, T.~Zhang, D.~Yu, and D.~Monteiro.
\newblock Ringtext: Dwell-free and hands-free text entry for mobile
  head-mounted displays using head motions.
\newblock {\em IEEE transactions on visualization and computer graphics},
  25(5):1991--2001, 2019.

\bibitem{zhang2010annotating}
B.~Zhang, Q.~Li, H.~Chao, B.~Chen, E.~Ofek, and Y.-Q. Xu.
\newblock Annotating and navigating tourist videos.
\newblock In {\em Proceedings of the 18th SIGSPATIAL International Conference
  on Advances in Geographic Information Systems}, pp. 260--269, 2010.

\bibitem{zhou2021partially}
Z.~Zhou, L.~Wang, and V.~Popescu.
\newblock A partially-sorted concentric layout for efficient label localization
  in augmented reality.
\newblock {\em IEEE Transactions on Visualization and Computer Graphics},
  27(11):4087--4096, 2021.

\end{thebibliography}
\end{document}